\documentclass[12pt]{article}

\usepackage{arxiv}

\pdfoutput=1

\usepackage[utf8]{inputenc} 
\usepackage[T1]{fontenc}    
\usepackage{hyperref}       
\usepackage{url}            
\usepackage{booktabs}       
\usepackage{amsfonts}       
\usepackage{nicefrac}       
\usepackage{microtype}      
\usepackage{dsfont}
\usepackage{amsmath}
\usepackage{xltabular}
\usepackage{soul}
\usepackage{cancel}
\usepackage{subfig}
\usepackage{floatrow}
\usepackage{listings}
\usepackage{hyperref}
\usepackage{graphicx}
\usepackage{caption}
\usepackage{fancyhdr}
\usepackage{tabularx}
\usepackage{lipsum}
\usepackage{xcolor}
\usepackage{multirow}
\usepackage[title]{appendix}

\title{
Dynamic portfolio optimization with \\
Inverse Covariance Clustering}

\author{
  Yuanrong Wang\\
  Department of Computer Science\\
  University College London\\
  Gower Street, London WC1E 6BT \\
  \texttt{yuanrong.wang@cs.ucl.ac.uk} \\
   \And
 Tomaso Aste \thanks{Corresponding Author} \\
  Department of Computer Science\\
  University College London\\
  Gower Street, London WC1E 6BT \\
  \texttt{t.aste@ucl.ac.uk} \\
}

\begin{document}
\maketitle
\begin{abstract}
Market conditions change continuously. 
However, in portfolio's investment strategies, it is hard to account for this intrinsic non-stationarity.
In this paper, we propose to address this issue by using the Inverse Covariance Clustering (ICC) method to identify inherent market states and then integrate such states into a dynamic portfolio optimization process.
Extensive experiments across three different markets, NASDAQ, FTSE and HS300, over a period of ten years, demonstrate the advantages of our proposed algorithm, termed Inverse Covariance Clustering-Portfolio Optimization (ICC-PO). 
The core of the ICC-PO methodology concerns the identification and clustering of market states from the analytics of past data and the forecasting of the future market state. It is therefore agnostic to the specific portfolio optimization method of choice. 
By applying the same portfolio optimization technique on a ICC temporal cluster, instead of the whole train period, we show that one can generate portfolios with substantially higher Sharpe Ratios, which are statistically more robust and resilient with great reductions in maximum loss in extreme situations. This is shown to be consistent across markets, periods, optimization methods and selection of portfolio assets. 
\end{abstract}


\section{Introduction} \label{Intro}
In the field of asset management, the problem of portfolio allocation has gained unprecedented popularity over the past few years.  Constructing a good portfolio combines the art and science of balancing between trade-offs and the aim to meet long-term financial goals. The simple core of any portfolio optimization is to assign optimal weights to each portfolio's component in order to minimize investment's risk and maximize the return. In 1952,  Markowitz \cite{Markowitz} demonstrated that, by assuming risk to be quantifiable by the variance of the portfolio's returns, the optimal weights which minimize portfolio's variance at a given average portfolio's return can be computed with a simple and exact formula. However, the Markowitz's theoretical maximum is attained only in-sample, on the train dataset, whereas off-sample, on the test set where investment is made, performances of the Markowitz's portfolio can be largely sub-optimal. 

Markowitz's modern portfolio theory is the foundation to modern quantitative asset management.
There are however two main limitation in the  Markowitz's assumptions. 
The first limitation concerns the use of the portfolio's variance as measure of risk. The variance (when defined) is indeed a measure for the width of the distribution but there are other properties that are better measure of risk (e.g. the value at risk) and might not be reducible to the variance when  the underlying probability distribution is not a location-scale.
The second limitation concerns the ability to estimate the (future) means and covariance of the asset's returns in the portfolio. 

After  Markowitz's  seminal work, many portfolio selection methodologies have been introduced to cure the first limitation concerning the reliance on variance for risk quantification and nowadays there are several well-established approaches that go well beyond the use of variance as sole risk measure \cite{Hult2012RiskAP}.
Furthermore, with the enormous development of machine learning optimization techniques there are presently virtually no limitations in constructing optimal portfolios based on any kind of risk measure \cite{KomSamo2016StochasticPT,Ban2018MachineLA,Paiva2019DecisionmakingFF}. 

Addressing the second limitation is harder. Indeed, normally, one does not have information from the future that would allow to set the future properties of the asset's multivariate distribution. 
Therefore, the reliance on  past observations and the assumption that they will significantly represent also the future, is hard to avoid.
Nonetheless, markets are not stationary, it is common knowledge that they cyclically pass through bull and bear states and occasionally deepen into crisis periods. For each of these periods the market prices' returns have different statistical properties and they are not describable by means of a unique multivariate probability distribution.  
This is especially relevant for factors that matter most to the management of portfolio risk. Indeed, crisis periods have distribution with fatter tails and they tend to be more asymmetrical with the left tail having larger probability for large losses than the right tail for equivalent gains. 
Portfolio constructions must take into account these differences and device different investment strategies for each market condition. This is indeed the ground basis for any dynamic asset allocation. However, such a wise allocation would imply the knowledge of the future market state and forecasting it from past observations is not an easy task.

In this paper, we provide an algorithm termed Inverse Covariance Clustering-Portfolio Optimization (ICC-PO) to address the non-stationarity problem, by identifying the inherent market states and forecast the most likely future state.
The Inverse Covariance Clustering (ICC) \cite{procacci1} is a novel temporal clustering method for market states clustering.
In this paper we propose to make use of this temporal clustering classification, constructing different optimal portfolios associated with two ICC market state clusters. 
The clusters are constructed in the in-sample training set (the past) and then are used separately to train the portfolio optimizer of choice which is then tested on an off-sample period following the training set (the future).
For the optimization we used two approached based on the 
classical Markowitz's approach but devised to have only positive weights (no short-sellings). They are the Sequential Least Square Quadratic Programming (SLS) and the Critical Line Algorithm (CLA). Let us note, that the ICC-PO approach allows the use of any optimization method of choice. 
We tested the approach with three extensive experiments with daily data, from 2010 to 2020, from three different markets: NASDAQ, FTSE and HS300. 
For each market, we selected 100 largest market capitalization constituent stocks and quantified the off-sample performances of portfolios constructed from in-sample training data using separately the two  ICC-market states. 
We demonstrate that the difference in returns and risks (computed on the testing set) between the two optimal portfolios, constructed from the two ICC-market states (on the training set), is very large with Sharpe Ratios that more than double and with very large differences in the likelihoods of large negative returns that can have up to three time smaller quantiles (i.e. value at risk). We provide a simple criteria to forecast the best performing out of sample market state which we named `State 0'.
Our results also show that sparsification of the inverse covariance matrix through information filtering networks  \cite{CompNet5,CompNet7} is improving the results, this is a confirmation of a previous result \cite{procacci2} extended however in this paper to a different dataset, different portfolio optimizers and different markets. 
The robustness of the method is tested by gathering statistics over 100 re-sampling of consecutive train-test sets randomly selected across the 10 years period 2010 to 2020. 
Furthermore, reliance on portfolio basket choices is tested by doing the same experiments with random selection of 100 stocks instead of the 100 most capitalized. 

The remaining of the paper is organized as follows: in Section 2, we review literature on market states clustering, dynamic portfolio optimization, and information filtering network applied in the experiments; in Section 3, we outline experimental methodologies and implementations. The final results are presented in Section 4 and discussed in Section 5.

\section {Background literature}  \label{lit}
\subsection{Mean-variance optimization}
Despite the unquetionable merits and pioneering status of Markowitz's mean-variance optimization (MVO) approach, there are some major assumptions, and several bad applications, that reduce its efficacy for practical implementations. 
Firstly, MVO assumes that asset returns follow a finite-variance distribution and higher moments are monotonic with variance. 
Many financial theories simplify this assumption adopting a normal distribution, and consequently the models that utilize such theories do not account for extreme market situations. Moreover, the variance of a normal distribution as a risk measure does not distinguish  between upside and downside moves in the market. 
Secondly, the MVO rely on the inversion of a covariance matrix and this operation makes the method  highly sensitive to estimation error especially when the covariance is estimated on a relatively short time-period and when such a past period is not representative of the future. 
Indeed, historical financial market data is never a good representation of the true underlying distribution as the observations are often partial. Furthermore, most MVO implementations are assuming market stationarity, which is that the mean and variance are assumed constant in each asset, while the correlation is static between assets. 
MVO is designed to avoid unsystematic risks by optimizing diversification. 
However, the systematic risks from market movements are not addressed by the MVO methodology and this is usually the most significant factor for investment decisions. 
Lastly, a single-period investment will almost never work in reality. 
A constant re-allocation is vital to respond to the rapidly changing environment. 

\subsection{Fat-tailed and asymmetric return distributions}
Normal distributions do not represent well the observed probability distribution of financial market's assets prices returns. 
Indeed, they instead have a larger number of small returns than what expected from normal statistics, but also a larger number of very large positive and negative returns of sizes that would be impossible with normal statistics \cite{Officer1972TheDO,Limpert2011ProblemsWU,Kchler1999StockRA}. They also have often asymmetric distributions with larger negative returns (losses) more likely than large positive ones (gains).
Several alternative probability distributions have been used in the literature,  namely, Student-t \cite{studentT0,studentT1,studentT2}, Laplace \cite{Laplace0,Laplace1,Laplace2} and Pareto-Levy \cite{Levy0,Levy1} distributions. 
In addition, alternative approaches to account for asymmetry have been taken into account, with early works by Markowitz himself which in 1959  \cite{Markowitz1959} employed semi-covariance (the covariance from negative returns only)  as a better risk measure to better describe the downside market moves. 
Furthermore, limited sample size is a critical contributing factor to estimation errors. Yet, simply extending the sample size introduces data from events happened far in the past which are likely to be less representative of present market conditions. Hence, methods ranging from shrinkage \cite{Ledoit2003HoneyIS}, to LASSO regularization \cite{Tibshirani1996RegressionSA,Friedman2008SparseIC}, and Monte Carlo based re-sampling \cite{MCresample1,MCresample2} have been used to reduce this issue. 

\subsection{Non-stationarity and dynamic portfolio allocation}
Assumptions regarding market stationarity and portfolio re-allocation are often considered together, since multi-period investment is proven to be an effective solution to mitigate the effect of Market turmoil. Several contributions have shown that dynamic re-allocation brings improvements in the resilience to market volatility with respect to the original single-period portfolio diversification methods \cite{diversification1,diversification2,diversification3}. 
Nonetheless, such methods still fail to address structured market movements. 
Indeed, accounting for such changes requires to forecast the future market state.
Further studies on market states has been proposed to model and predict the intrinsic properties of these dynamics, and two main streams are discussed below. The first one uses Markov decision process to model the transition probability between different market regimes. Currently, Hidden Markov Model (HMM) has demonstrated great efficiency and validity \cite{MarketStatesPO3}. However, it often encounters problems mainly associated with the curse of dimensionality, as the dimensionality of hidden states is linear to the number of assets considered \cite{MarketStatesPO1,MarketStatesPO2}. On the other stream, researchers believe that market comprises mixed multivariate distributions, and each state effectively corresponds to a distribution. Hence, temporal clustering methods such as Gaussian Mixture \cite{GMM1,GMM2,GMM3}, K-Nearest Neighbors (KNN) \cite{KNN1,KNN2,KNN3} have been applied for this purpose. Then, portfolios can be re-adjusted according to the predicted state with a selected re-allocation period. Yet, these methods often based on strong assumptions and they are not originally designed for time-series, which results issues e.g., Gaussian Mixture assumes Gaussian nature in all the base distributions, and KNN overlooks temporal consistency between single data point. This is also to some extent the approach of the decision-theoretic Bayesian methods \cite{Bayesian1,Bayesian2,Bayesian3}, such as the Black-Litterman model \cite{BL1,BL2}, which includes in the optimization a Bayesian prior on the future state. The ICC-PO approach introduced in the present paper is in the same line of temporal clustering methods just mentioned. However, in our case the temporal clustering is the ICC method and we make no use of Gaussian Mixtures.

The traditional Markowitz model optimizes on a single-period only, and it relies heavily on the assumption of constant asset mean vectors and covariance matrix. Therefore, this static and long investment horizon is inadequate in a dynamic market place. Yet, the mean-variance criteria inspires the development in multi-period dynamic portfolio construction. The dynamic portfolio optimization field currently follows two main streams. A discrete-time model was proposed by Samuelson in 1969 \cite{discretePO} and developed since by Hakansson, Grauer and others \cite{discretePO1, discretePO2, discretePO3}. It separates an investment horizon into discrete periods, and the portfolio can be reallocated at the end of each period. In contrast, a continuous-time model was introduced by Merton \cite{continuousPO} in the same year, and together with further studies described the continuous rebalancing of securities for a fixed planning horizon \cite{continuousPO1, continuousPO2, continuousPO3}.

The two alternative assumptions that are often made in dynamic portfolio optimization problems are market completeness and investment horizon. A complete market is an approximation to the real market where friction, transaction costs and asset liquidity exist, and dynamic portfolio has to consider those real world factors \cite{competeMarket1,competeMarket2,competeMarket3,competeMarket4}. A more ideal scenario is instead the incomplete market where some conditions are waived so that research can only focus on dynamic asset selection process and ignore some practical issues \cite{IncompeteMarket1,IncompeteMarket2,IncompeteMarket3,IncompeteMarket4}. Similarly, infinite horizon is a naive assumption to finite horizon where investors will withdraw investment with an exit time. The earlier pioneers \cite{discretePO1,continuousPO1,competeMarket2} in this field, such as Samuelson \cite{discretePO}, Merton \cite{continuousPO,continuousPO2}, began with the infinite horizon assumption, while later researchers in the 90s and the beginning of the millennium \cite{finiteHorizon3,finiteHorizon4, competeMarket4,IncompeteMarket4} led by He \& Pearson \cite{finiteHorizon1} and Karatzas et.al \cite{finiteHorizon2} started to introduce the finite horizon into the problem.

\subsection{Market States Clustering}

After the initial pitfall of Markov-Model-based methods \cite{MarketStatesPO1,MarketStatesPO2,MarketStatesPO3}, mainly due to the curse of dimensionality, literature has started to look for alternative methods to cluster similar temporal data points into a same group based on certain comparison criteria. Such temporal clustering methods can mostly be divided into two approaches: subsequent clustering and point clustering. Subsequent clustering uses a sliding window to capture a period of data points and analyze for recurrent patterns \cite{SubsequenceC0,SubsequenceC1}. The four main methods of subsequent clustering are: (i) hierarchical \cite{HierachicalC1,HierachicalC2,HierachicalC3}; (ii) partitioning \cite{PartitioningC1,PartitioningC2}; (iii) density-based \cite{DensityC1,DensityC2,DensityC3}and; (iv) pattern discovery \cite{PDC1,PDC2,PDC3}.
These method have all shown applicability to financial data analysis and portfolio construction. 
An alternative approach is point clustering that, instead of measuring spatial similarity between two slices of time-series, it looks at each temporal point individually, and assigns this multivariate observation to an appropriate cluster based on  distance metrics \cite{PointC1,PointC2,PointC3}. Hence, in point clustering, the choice of distance is  core. 
In macroeconomics, the market states are not the representation of solely upward or downward trends of the market, but also the relative dynamics of equity prices, which naturally makes correlations a convenient choice of collective dynamics. A stationary correlation structure was assumed as the common approach in the industry in the 90s \cite{BL2,statCorr1}, which was, however, later shown to be overly presumptive \cite{nostatCorr1,nostatCorr2,nostatCorr3}. Consequently, research has been devoted to study time-varying correlations. Models, such as Generalized Autoregressive Conditional Heteroskedasticity (GARCH) \cite{msClustering1} and the Dynamic Conditional Correlation (DCC) \cite{msClustering2} have been proposed for simulating and predicting this dynamical correlation. However, most of these models suffer from the curse of dimensionality and can only be applied to a limited number of assets, as numbers of parameters increases super-linearly with the number of variables.

In 2017, Hallac et al. proposed the Toeplitz Inverse Covariance Clustering (TICC) \cite{msClustering3} algorithm, originally devised for electric vehicles action sensor. It classifies states based on the likelihood measures of short subsequences of observations and corresponding sparse precision matrix. After clustering, the precision matrix of each state is estimated under a Toeplitz constraint. Inspired by TICC, Procacci and Aste in 2020  \cite{procacci1} proposed a closed related methodology names Inverse Covariance Clustering (ICC). This approach provides a point clustering of observations also enforcing temporal consistency by penalizing switching between states. The ICC method also uses sparse precision matrices but sparsification is attained via information filtering networks (see next Subsection). One main advantage of ICC, compared to TICC, is its flexibility in the selection of similarity measures. It was also stated in their original paper that different clustering distances separate market states differently. For example, likelihood distance distinguishes better with pre- and post-crisis period, Euclidean distance discriminates well between bull and bear states, and Mahalanobis distance is a mixture of the above.

\subsection{Information Filtering Networks}
Many computational methods employ sparse approximation techniques to estimate the inverse covariance matrix. 
The sparsification is effective because the least significant components in a covariance matrix are often largely prone to small changes and can lead to instability. 
Sparsified models filters out these insignificant components, and thus improve the model resilience to noise. 
A widely used approach for inverse covariance sparsification is GLASSO \cite{CompNet9} that uses $L_1$ norm regulatization.
An alternative approach that uses information filtering networks was shown to deliver better results with lower computational burden and larger interpretability \cite{CompNet7}.
Information filtering network analysis of complex system data has advanced significantly in the past few years. Its aim is to model interactions in a complex system as a network structure of elements (vertices) and interactions (edges). The first and best know approach is the Minimum Spanning Tree (MST) that was firstly introduced by Boruvka in 1926 \cite{nevsetvril2001otakar} and it can be solved exactly (see \cite{CompNet10} and  \cite{CompNet11} for two common approaches). 
The MST reduces the structure to a connected tree which retains the larger correlations.
To better extract useful information, Tumminello et al. \cite{CompNet3} and Aste and Di Matteo \cite{CompNet4} introduced the use of planar graphs in the Planar Maximally Filtered Graph (PMFG) algorithm. Recent studies have extended the approach to chordal graphs of flexible sparsity \cite{CompNet5, CompNet6}. Research fields ranging from finance \cite{CompNet7} to neural systems \cite{CompNet8} have applied this approach as a powerful tool to understand high dimensional dependency and construct a sparse representation. 
It was shown that, for chordal information filtering networks, such as the Triangulated Maximally Filtered Graph (TMFG) \cite{CompNet5}, one can obtain a sparse precision matrix that is positively definite and has the structure of the network paving the way for a proper $L_0$-norm topological regularization \cite{aste2020topological}. This approach has been proved to be computationally more efficient and stable than GLASSO \cite{CompNet9}, especially when few data points are available \cite{CompNet7, CompNet4}.

 \subsection{Sparse inverse covariance for portfolio construction and market state prediction}
The application of sparse inverse covariance for portfolio construction has been recently introduced in the literature \cite{Millington2017RobustPR,Yuan2020ImprovedLD,Lee2020OptimalPU}.
The general approach has been to make use of $L_1$-norm regularization via GLASSO \cite{CompNet9}.
In a recent paper the sparsification methodology via information filtering networks, was applied to the identification of inherent market states via ICC \cite{procacci1}. 
In a following paper, the sparsification with TMFG information filtering networks was applied to the portfolio construction problem revealing several advantages with respect to traditional mean-variance methods \cite{procacci2}.

\section{Methodologies}
In the present paper we combine ICC clustering with market state forecasting to be used for portfolio optimization. 
Let us list in this Section the main methods we use in our approach.
 
\subsection{Inverse covariance temporal clustering for portfolio optimization (ICC-PO)} 
Let's consider a set of $n$ assets with $\mathbf r_t \in \mathbb R^{1\times n}$ the vector of returns at time $t$. The corresponding vector of their expected values is $\boldsymbol \mu = \mathbb E(\mathbf r_t) \in \mathbb R^{1\times n}$ and their covariance matrix is $\boldsymbol{\Sigma} = \mathbb E((\mathbf r_t-\boldsymbol \mu)^\top (\mathbf r_t-\boldsymbol \mu))\in \mathbb R^{n\times n}$.
The ICC clustering method depends on the choice of a gain function, $G_{t,k}$, which is a measure  which qualifies the gain when the time $t$ returns, $\mathbf r_t$, are associated whith cluster $k$.
Indeed,  the ICC approach gathers together in cluster $k$ observations that have the largest gain in such a cluster with respect to any other cluster: $G_{t,k} > G_{t,h}$ for all $h\not= k$.  
For instance, in \cite{procacci1} it was used
\begin{equation}\label{eq:likelihodEuclidean}
    G^{Eu}_{t,k} =   -(\mathbf {r}_t-\hat {\boldsymbol \mu}_k)(\mathbf {r}_t-\hat {\boldsymbol \mu}_k)^\top
\end{equation}
where $\hat {\boldsymbol \mu}_k$ is the sample mean return computed form the observations in cluster $k$. This gain is  minus the square of the euclidean distance between the observation and the centroid of cluster $k$.
A distance associated with the likelihood for multivariate normal distributions is instead
\begin{equation}\label{eq:likelihodN}
    G^{No}_{t,k} =  \frac{1}{2} \ln | \hat {\boldsymbol \Sigma}_k^{-1} | - n \frac{d^2_{t,k}}{2},
\end{equation}
with
\begin{equation}\label{eq:Maha}
d^2_{t,k} = (\mathbf{r}_t-\hat {\boldsymbol \mu}_k)^\top \hat {\boldsymbol \Sigma}_k^{-1}(\mathbf{r}_t-\hat {\boldsymbol \mu}_k)
\end{equation}
the Mahalanobis distance where $\hat {\boldsymbol \Sigma}_k$ is the sample covariance computed form the observations in cluster $k$.
While for the multivariate Student-t one has
\begin{equation}\label{eq:likelihodSt}
    G^{St}_{t,k} =  \frac{1}{2} \ln |\hat {\boldsymbol \Sigma}_k^{-1} | - \frac{\nu+n}{2}\ln (1+\frac{d^2_{t,k}}{\nu})
\end{equation}
where, in this case, $\boldsymbol \Sigma_k^{-1}$ is the scale matrix, which is $1-2/\nu$ times the sample covariance when $\nu>2$ and the covariance is defined (note that also  $\hat {\boldsymbol \Sigma}_k$ in the expression for $d^2_{t,k}$ is the scale matrix, in this case).

We extensively tested all these gain functions observing that $G^{Eu}_{t,k}$ is particularly efficient is selecting clusters with prevalence of positive or negative returns but it is performing poorly in the portfolio optimization problem. The normal and Student-t likelihood related gains have similar performances, but $G^{St}_{t,k}$ turns out to be in average superior and we adopted it for the experiments we present in this paper.
We also tested an hybrid distance $G_{t,k} =  c_1 \ln | \hat {\boldsymbol \Sigma}_k^{-1}  | - c_2 {d}^2_{t,k}$ with the two arbitrary constants, $c_1$ and $c_2$, that allow to gauge between the effects of the determinant of the covariance (an entropic term) and the Mahalanobis distance term. The measure of the determinant of a covariance is an equivalent estimation of the differential entropy of the multivariate system, while Mahalanobis distance measures distance between points and distributions.

ICC approach uses sparse inverse covariance that was shown to improve considerably results over the full covariance. As sparsification technique we used the sparse inverse constructed with TMFG information filtering graphs \cite{CompNet5} using the local-global (LoGo) inversion procedure described in \cite{CompNet7}, where the elements of the inverse are computed by inverting local sample covariance matrices from only four variables at the time and adding them up. 
The result is a sparse inverse covariance with $3n-6$ non-zero entries in the upper diagonal (instead of $n(n-1)/2$ in the full matrix). 
Such a matrix is  positively definited, if the number of observations is larger than four, independently on the size of the whole matrix ($n \times n$).
Sparse portfolios are  simply obtained by applying a portfolio optimization method (see next subsection) with a sparse inverse covariance instead of a full covariance as input. 

A final key element of the ICC methodology is the temporal consistency of the cluster that is imposed by penalizing frequent switches between clusters. In this paper the penalizer parameter $\gamma$ is estimated in the train set through a grid search so that the average cluster persistence is of a given length (30 days in this paper).

The assignment of the temporal instance $t$ to a cluster number, $k_t$, is performed iteratively starting from an initial random cluster assignment. Specifically we evaluate the penalized gain
\begin{equation}\label{eq:PenGain}
    \tilde G_{t,k_t} =  G_{t,k_t} - \gamma \delta_{k_{t-1},k_t},
\end{equation}
and assign observation $t$ to the cluster with largest penalized gain. In the previous expression, $\delta_{k_{t-1},k_t}$ is the Kronecker delta returning one if $k_{t-1}=k_t$ and zero otherwise. After the assignment of the time-$t$ observation to a given cluster $k_t$, all cluster parameters (means and covariances) are recomputed with the new cluster assignments. 

We then performed a mean variance portfolio optimization method independently for each ICC state. Obtaining optimal weights associated with each temporal cluster. To apply effectively such optimized weights to the portfolio problem we have to forecast the state that is most likely to be predominant in the future test set where the investment is performed. For this purpose we made use of the short term persistence of such states and we assigned as most likely future state the one that is predominant in the last part of the train set. In this paper we consider two clusters only. 

\subsection{Portfolio optimization methods}
Our proposed methodology is made of three main stages.
First, we use ICC for  temporal clustering the train dataset into two market states. Second, we  forecast  which of the two states will be predominant in the future, test dataset, where the investment is made. Third, we perform portfolio optimization using train data from the forecasted predominant ICC state. 
Our approach is, to large extent, agnostic to the kind of optimization adopted. In this paper, for the experiments, we used two, mean-variance optimizations methods:  1. the Sequential Least Square Quadratic Programming approach and; 2. the Critical Line Algorithm method. Let us briefly recall the basic elements of these two portfolio optimization methods. 

For the experiments in this paper Markowitz's optimal weights can be computed with the python package 'Numpy' for direct matrix multiplication. The exact solution is shown in Appendix \ref{appendixF}. In the literature, this solution is referred to as `unconstrained' because, beside the normalization and average conditions, the weights have no other constraints. On the other hand, in some practical cases, one might want to add further conditions to the weights. For instance, many real world situations do not allow short selling, which hence makes necessary to impose only positive weights in the range $w_i \in [0,1]$. This constrained optimization problem cannot be any longer analytically solved and numerical optimization methods must be adopted. 

Two numerical optimization methods have been adopted in the experiment. The {\bf sequential least square quadratic programming (SLS)} \cite{Quadratic_optimisation0,SLSQP1,SLSQP2} is considered to be one of the most efficient computational method to solve general nonlinear constrained optimization problems. Jackson et al. and Cesarone et al. demonstrate its effectiveness in finance \cite{Quadratic_optimisation1,Quadratic_optimisation2}. There is an easy-to-use package implemented in Python's {\fontfamily{cmr}\selectfont \textbf{SciPy.optimize}} library \cite{scipy_sls} which we applied in our experiments. The {\bf Critical Line Algorithm (CLA)} is an efficient alternative to the quadratic optimizer for mean-variance model, as it is specifically designed for inequality portfolio optimization. It was already originally introduced in the Markowitz Portfolio Selection paper \cite{Markowitz}, and its computational implementation has become increasingly popular \cite{CLA2,CLA3}. CLA also solves constrained problems with conditions in inequalities, but unlike SLS, it divides a constrained problem into series of unconstrained sub-problems. In our experiment, to compute CLA optimization for portfolio selection, we leveraged the implementation from the open-source {\fontfamily{cmr}\selectfont \textbf{portfoliolab}} Python library from Hudson and Thames \cite{CLA1}. A key drawback of CLA is called the Curse of Markowtitz, which is that a small change can lead to a very unstable inverse covariance matrix calculation. Our employment of sparse inverse covariance matrices via information filtering network produces more robust results that are more resilient to noise produced by small changes, and the overall model deliver better performances with respect to the model with full inverse covarainces. Mathematical and algorithmic details of SLS and CLA are included in included in Appendix \ref{appendixF} for reference.

In summary, these two portfolio optimization methodologies output optimal portfolio  weights $\mathbf W$ from an input constituted of: (i)  a set of observations $\mathbf r_t$; (ii)  a vector of mean returns  $\boldsymbol \mu$; (iii)  a covariance $\boldsymbol \Sigma$. As we shall see shortly, in our implementation these inputs are provided in various combinations including selecting from ICC states and sparsifying.

\section{Implementation} \label{imp}

\subsection{Data}
We carried out several experiments using historical financial time-series data from three major capital markets: NASDAQ, FTSE and HS300. We selected 100 stocks from each of these three markets during the trading period between 01/01/2010 and 01/01/2020. 
For each stock, we calculated the daily log-return, $r_i(t)=\log(P_i(t))-\log(P_i(t-1))$, using closing prices.
For the 100 stocks, in the main paper, we selected the largest market capitalization constituents but in the appendix we repeat the experiments with random selection obtaining comparable results. 

\subsection{Experiments}

The optimal portfolio weights are obtained from the data in the train set  and performances are measured over the test set where portfolio weights are left unchanged. 
As performance indicators, we compute portfolio return, portfolio standard deviation (i.e. volatility) and Sharpe ratio over the investment horizon (test period). 
We report the annualized value of these quantities, estimated as the daily values multiplied by $\sqrt{252}$.
For statistical robustness, for each market, we compute the above portfolio performance indicators over 100 randomly chosen consecutive train-test periods within the ten years dataset. Results are reported for the mean performances and the 5\%-95\% quantile ranges over such re-sampling.

\begin{figure}[h]
        \centering
        \subfloat[Average Sharpe Ratio for 4 States]{\includegraphics[width=\linewidth]{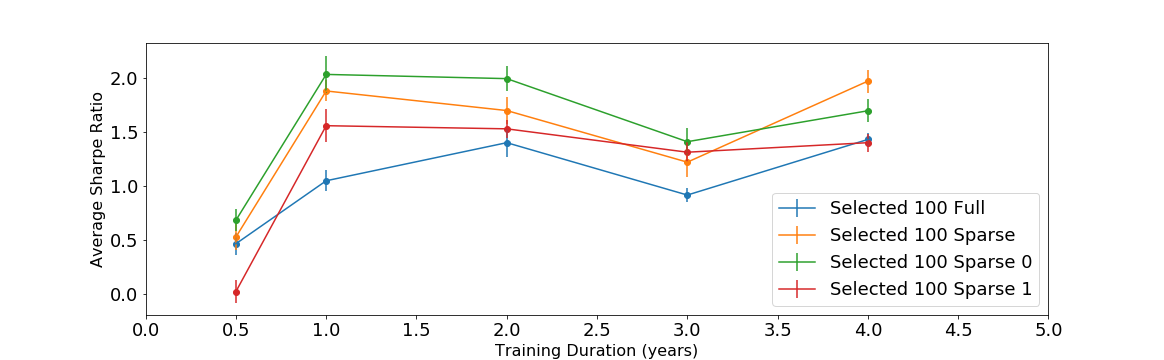}}
        \newline
        \subfloat[The relative Sharpe Ratio between Sparse 0 and Full]{\includegraphics[width=\linewidth]{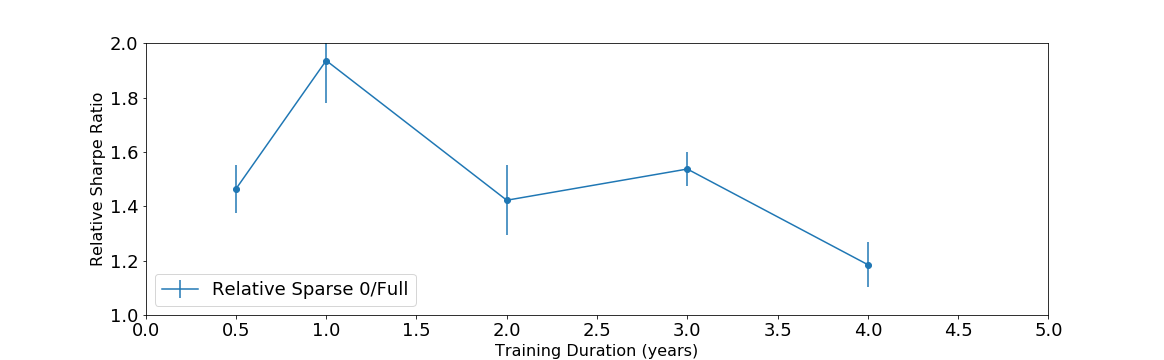}}
        \vspace{0.5pt}
        \caption{
        Sharpe Ratio for portfolios with 100 largest market capitalization constituent stocks of NASDAQ Composite optimized using different training set durations. The top subplot reports the average Sharpe Ratios ($SR$) with error bars reporting 1 standard deviation, for Full, Sparse, Sparse 0 and 1, statistics is on 100 training-testing periods chosen at random within the 10 years dataset.
        The bottom subplot report instead the relative Sharpe Ratios between Sparse 0 and Full, $SR_{Sparse 0}/SR_{Full}$.  }
        \label{fig:Selected training duration}
    \end{figure}

The test set length (investment horizon) was established at 30 days which is a reasonable value for practical applications, we however also report in appendix results for horizons of 10, 20 and 100 days finding consistent results.
The train set length was established by performing experiments with train sets of  $L= 0.5, 1 ,2 ,3 ,4$ years.
Figure \ref{fig:Selected training duration}  reports the annualized average Sharpe Ratio computed on the test set as function of the  train set length. 
One can observe from the top figure that the lengths between one and two years yields to consistent good performances. We adopted the period of 2 years as optimal compromise between statistical robustness and best performances.

In the experiments, we first  compute, on the training set, the ICC time clusters assuming two states and Student-t log-likelihood, Eq.\ref{eq:likelihodSt} as gain function. 
The choice of two states is for the seek of simplicity, we tested also 3 states obtaining inferior but comparable results.  
We verified that Student-t likelihood is best performing among the tested gain functions, in Appendix we report results also for Normal log-likelihoods (Eq.\ref{eq:likelihodN}).
The switching penalty parameter $\gamma$ in ICC was set so that the average cluster size is around 30 days, i.e. consistent with the 30-day investment horizon. This selection of average cluster size and investment horizon is a somehow arbitrary choice based on the effective threshold of the portfolio performance measured by the Sharpe Ratio. 
We then labeled `Sparse 0' the state that is most abundant among the last 20 days of observations at the end of the train period. Conversely, we labeled `Sparse 1' the other. The term Sparse is used to indicate that this portfolio uses sparse inverse covariance.
To set such a `prevalence period' of 20 days we first performed a grid search over the combination of training duration $L= 0.5, 1 ,2 ,3 ,4$ years, and using prevalence periods of $10, 20 ,30, 100, L/2, L $ days. This search confirmed that small values of prevalence periods, of $10, 20, 30$ days, provide better results than larger prevalence periods. We therefore set a prevalence period of $20$ days as it provides the most consistent results across the grid search and it is also consistent with the length of the test set. 

The bottom plot in Figure \ref{fig:Selected training duration} reveals that Sparse 0 has consistent better performances over Full with, best results for training periods of one year. 
Let us notice that, having a ICC average cluster size of 30 days, it makes hard to cluster well a small training period of six months, and often unbalanced clusters where one cluster dominates the period are obtained. 
On the contrary, a large training duration (4 years) makes the model prone to unnecessary patterns and noise, and in turn reduces  performance. 
Thus we chose 1 year as best compromize for the length of the training set. In Appendix \ref{appendixA} and \ref{appendixC} we see that similar results are obtained for the other two markets (FTSE, HS300).

We  compute optimal portfolios using the two (SLS and CLA) optimization methods.
We trained each optimization method both on the whole train dataset and, separately, on the two Sparse 0 and Sparse 1 states.
We used the sample means for each of the respective sets and either the `full' sample covariances (Pearson's estimate) or the `sparse' sample covariances  (TMFG-LoGo estimate  \cite{CompNet5,CompNet7}). 
Therefore, for each optimization method we have four optimized portfolios: 
two  computed on the whole training set and with full or sparse inverse covariance (named `Full' and 'Sparse' respectively);  two computed on the two ICC market states and with sparse inverse covariance (named `Sparse 0' and `Sparse 1').
For benchmarking, these portfolios are also compared to a  portfolio with equal weights, $w_i=1/n$ named `Naive'. 
Overall, we have therefore $4 \times 3$ plus 1 differently optimized portfolios that are recomputed 100 times over randomly sampled time-periods.
Such optimized portfolio weights are applied, for each of the three markets, to 100 most capitalized stocks. In appendix we repeat the experiments for randomly selected  stocks.

\section{Results}
    

\subsection{Log-likelihood}\label{Likelihood_results}
\begin{figure}[h]
        \centering
        \includegraphics[width=\linewidth]{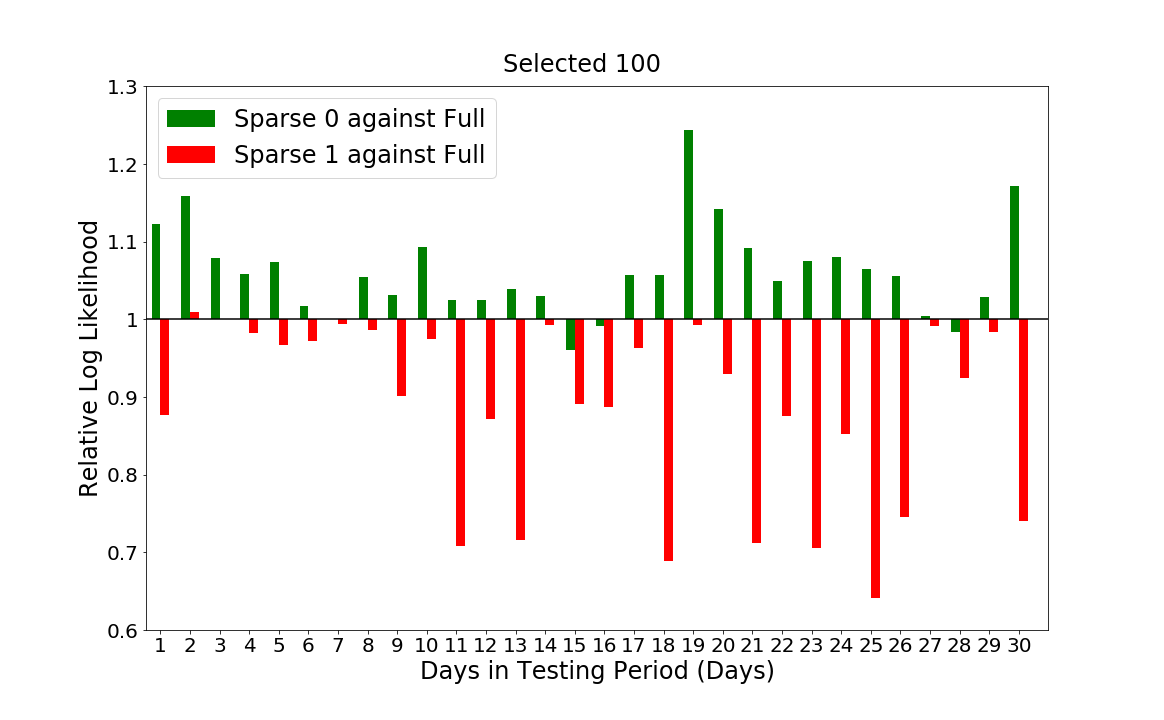}
        
        \vspace{0.5pt}
        \caption[Likelihood plots]{
 Student-t log-likelihood for 100 largest market capitalization constituent stocks of NASDAQ Composite v.s. number of days in the test period after training. 
Each bar represents the average gain of the Sparse 0 (green) or 1 (red) with respect to the Full in each day. Averages are over 100 re-samplings.}
         \label{fig:Selected_Likelihood_plots}
\end{figure}

We computed the daily Student-t log-likelihood, using Eq.\eqref{eq:likelihodSt}, for each of the 30-day investment horizon. 
Figure \ref{fig:Selected_Likelihood_plots}, reports for the averages of the differences for each day between the log-likelihood of Sparse 0 and full and also between Sparse 1 and full. The average is taken over the 100 random re-sampling.

 Figure \ref{fig:Selected_Likelihood_plots}, shows mostly positive gains for Sparse 0 indicating that, for most days across the investment horizon, it has larger log-likelihoods than Full. 
 Sparse 1 gain instead reveals mostly negative results against full. 
 This therefore indicates that while Sparse 0 is, in average, a better model to describe the multivariate nature of the log-returns in the test set with respect to Full; instead, Sparse 1 is in average worst. 
Since both Sparse 0 and 1 were sparsified using TMFG, the difference between them must therefore  be a consequence of clustering. One might note that, even though some Sparse 1 log-likelihoods gains are in the positive domain, they anyway have smaller magnitudes than their Sparse 0 counterparts. This result clearly shows the effectiveness and importance of considering market states.

Let us note that the two ICC clusters gather together observations that maximize in-sample log-likelihood in the respective clustered periods.
The fact that these models (i.e. in-sample means and covariance) from these clusters still correspond to different log-likelihood performances in the off-sample test set indicates that the states are still relevant off-sample.
Further, the better off-sample performances of Sparse 0 state indicates predictability; i.e. if one state outperforms another during the training period, it will remain better performing throughout the test period. Furthermore, to illustrate the universality of the log-likelihood results, we have included similar Student-t log-likelihood (Appendix \ref{appendixA}) as well as Normal log-likelihood (Appendix \ref{appendixC}) plots for random 100 stocks selections for all three major indices (NASDAQ, FTSE, HS300), where similar patterns are observed. 

\subsection{Portfolio Performance} \label{portfolio_results}
We tested portfolio performances over 30-day investment horizon for 100 largest market capitalization constituent stocks of NASDAQ Composite computed with the four portfolio optimization methods, SLS and CLA and using as inputs Full, Sparse, Sparse 0 and Sparse 1. We also report the $1/n$ Naive construction for benchmark. Tables \ref{Tab:ST_Selected30} and \ref{Tab:Normal_Selected30} reports portfolio performances for the combination of portfolio constructions (column `Solver') and inputs (column `State').  
Performances are quantified in terms of annualized portfolio return, annualized portfolio standard deviation (volatility) and the annualized Sharpe ratio over 30-day investment horizon. We report the 5\% and 95\% quantiles and the means computed from the 100 random resampling of  consecutive training-investment periods chosen at random within the 10 years dataset.  
The maximum returns and Sharpe Ratios, as well as the minimum volatility, are highlighted in bold. Thus showing the best performer in each market-solver combination. In addition, we highlight the minimum 5th percentile returns to depict the state suffering the least loss, and the maximum 95th percentile volatility to showcase the most stable state in extreme market situations.

\begin{table}[]
    \centering
\begin{tabular}{cc|c p{2cm}|cp{2cm}|cp{2cm}}
\toprule
 Solver &     State &  Return (\%) & (5,95)th percentile &  Volatility (\%) & (5,95)th percentile &  Sharpe & (5,95)th percentile \\
\midrule
       &    $\frac{1}{n}$ Naive &   14.46 &                    (-36,55) &       17.4 &                     (14,28) &   1.536 &                  (-1.4,4.3) \\
\midrule
    SLS &      Full &   22.71 &                    (-28,96) &       19.5 &                     (14,28) &   1.627 &                  (-1.4,5.4) \\
    SLS &    Sparse &   21.81 &                    (-23,74) &       17.5 &                     (14,26) &   1.764 &                  (-1.0,6.2) \\
    SLS &  Sparse 0 &   \textbf{29.04} &                     (\textbf{-6},66) &       \textbf{16.0} &                     (12,\textbf{25}) &   \textbf{2.478} &                  (-0.3,6.9) \\
    SLS &  Sparse 1 &    5.35 &                    (-49,57) &       19.8 &                     (14,34) &   0.978 &                  (-2.3,4.6) \\
 \midrule
    CLA &      Full &   21.97 &                    (-69,97) &       19.5 &                     (14,31) &   1.541 &                  (-2.1,6.5) \\
    CLA &    Sparse &   22.27 &                    (-32,85) &       17.0 &                     (12,27) &   1.758 &                  (-1.9,6.5) \\
    CLA &  Sparse 0 &   \textbf{28.73} &                    (\textbf{-27},76) &       \textbf{15.8} &                     (11,\textbf{26}) &  \textbf{2.372} &                  (-1.5,7.6) \\
    CLA &  Sparse 1 &   12.48 &                    (-57,86) &       18.7 &                     (12,32) &   0.964 &                  (-2.9,6.6) \\
\bottomrule
  \\
    \end{tabular}
    \vspace{0.5pt}
    \caption{
 Portfolio performances obtained by using Student-t log-likelihood for ICC clustering.
 We report annualized  return, annualized volatility and annualized Sharpe Ratio computed on 30 days investment period after the 1 year training set.
The values are averages and 5th and 95th percentiles computed over 30-day investment horizon from obtained from 100  re-sampling of  consecutive training-investment periods chosen at random within the 10 years dataset. 
 The underlying assets are 100 largest market capitalization constituent stocks of NASDAQ Composite. Highlight  in bold are return, volatility and Sharpe Ratio indicating the optimal state in each market solver combination, while highlights in 5th return and 95th volatility showcase the extreme behaviours (excluding the state Market). The state $1/n$ Naive is the equally weighted un-optimised portfolio and it is reported as benchmark. 
 }
    \label{Tab:ST_Selected30}
\end{table}

\begin{table}[]
    \centering
\begin{tabular}{cc|c p{2cm}|cp{2cm}|cp{2cm}}
\toprule
 Solver &     State &  Return (\%) & (5,95)th percentile &  Volatility (\%) & (5,95)th percentile &  Sharpe & (5,95)th percentile \\
    \midrule
       &    $\frac{1}{n}$ Naive &   14.46 &                    (-36,55) &       17.4 &                     (14,28) &   1.536 &                  (-1.4,4.3) \\
\midrule
   SLS &      Full &   22.98 &                       (-28,96) &        19.3 &                        (14,28) &   1.667 &                  (-1.4,5.4) \\
   SLS &    Sparse &   21.96 &                       (-23,74) &        17.3 &                        (14,26) &   1.787 &                  (-1.0,6.2) \\
   SLS &  Sparse 0 &   \textbf{29.00} &                       (\textbf{-14},66) &        \textbf{15.9} &                        (12,\textbf{23}) &   \textbf{2.260} &                  (-0.8,4.6) \\
   SLS &  Sparse 1 &    6.84 &                       (-43,63) &        19.5 &                        (14,30) &   0.845 &                  (-1.8,4.6) \\
   \midrule
   CLA &      Full &   20.63 &                       (-76,97) &        19.5 &                        (14,31) &   1.456 &                  (-3.0,6.5) \\
   CLA &    Sparse &   21.15 &                       (-53,85) &        17.0 &                        (12,\textbf{27}) &   1.678 &                  (-2.1,6.5) \\
   CLA &  Sparse 0 &   \textbf{27.08} &                       (\textbf{-14},79) &        \textbf{15.6} &                        (10,30) &   \textbf{2.175} &                  (-0.7,6.6) \\
   CLA &  Sparse 1 &   11.54 &                       (-69,77) &        18.6 &                        (14,36) &   1.028 &                  (-2.6,5.6) \\
\bottomrule
  \\
    \end{tabular}
    \vspace{0.5pt}
    \caption{Portfolio performances obtained by using Normal log-likelihood for ICC clustering.
 We report annualized  return, annualized volatility and annualized Sharpe Ratio computed on 30 days investment period after the 1 year training set.
The values are averages and 5th and 95th percentiles computed over 30-day investment horizon from obtained from 100  re-sampling of  consecutive training-investment periods chosen at random within the 10 years dataset. 
 The underlying assets are 100 largest market capitalization constituent stocks of NASDAQ Composite. Highlight  in bold are return, volatility and Sharpe Ratio indicating the optimal state in each market solver combination, while highlights in 5th return and 95th volatility showcase the extreme behaviours (excluding the state Market). The state $1/n$ Naive is the equally weighted un-optimised portfolio and it is reported as benchmark.
    }
    \label{Tab:Normal_Selected30}
\end{table}

From the mean values reported in Tables \ref{Tab:ST_Selected30}  (Student-t log-likelihood), we observe  that Sparse 0 outperforms Full, and this supremacy dominates for the two solvers. More specifically, Sparse 0 is on average $29.3\%$, $19.5\%$ and $53.1\%$ better in return, volatility and Sharpe Ratio than Full across all two solvers. 
We observe instead that Sparse 1 is considerably worst than Full indicating therefore that the significant gain of State 0  come  from filtering out the `disadvantageous' Sparse 1 state rather than sparsification.
This is indeed confirmed by the small observed gains of Sparse over Full.
these results are confirmed by the analysis of the 5th and 95th quantiles where we notice that Sparse 0 consistently achieves the least minimum extreme loss and the least maximum extreme risk. Specifically, Sparse 0 on average loses $66.0\%$ less and is $13.6\%$ less volatile than Full on 5th percentile return and 95th percentile volatility respectively. In other words, the integrated clustering portfolio optimization algorithm, ICC-PO, that we proposed can boost returns with less risk than the traditional benchmark, as well as provide extra resilience in extreme market situations. 

To test the sensitivity of this method to the specific ICC clustering gain function, we performed the same analysis using normal log-likelihood gain function for ICC clustering. 
Results are reported in Table \ref{Tab:Normal_Selected30}. 
Consistently with the previous results, we observe $28.6\%$, $18.8\%$ and $42.0\%$ improvements in return, volatility and Sharpe Ratio, with $73.1\%$ and $10.2\%$ gains in 5th percentile return and 95th percentile volatility. The comparison illustrates a $11.1\%$ Sharpe Ratio improvement in Student-t log-likelihood and a $7.1\%$ 5th percentile return advance in Normal log-likelihood. In other words, Student-t is a better model to the market and boosts portfolio performance. However, Normal log-likelihood generates a higher resilience to extreme loss. Since the average gain in return and volatility are similar in the two cases, the performance difference should mainly come from general upward-shifted ranges in the Student-t Sharpe Ratio.

Similar tables of optimization results using 10, 20, 30 and 100-day investment horizons, can be found in Appendix \ref{appendixB} for Student-t log-likelihood and \ref{appendixD} for Normal log-likelihood. 
These experiments were carried over randomly selected 100 stocks baskets (instead of the 100 most capitalized ones); the set of 100 random stocks was re-chosen for each of the 100 re-sampling.
Most of the general patterns found earlier still hold regardless of the length of the testing period and underlying assets. The relative difference, namely, the gain between Sparse 0 and Full remains roughly the same. This consistency further confirms the generality of our ICC-PO model.
In this case we report only the percentiles of the performance measures because being re-sampled on different constituents, mean values might be misleading. 

\section{Discussion} \label{diss}

The results presented in Section \ref{Likelihood_results} quantitatively demonstrate an effective gain in log-likelihood after applying temporal ICC clustering and computing the optimized sparse portfolio associated to the most persistent ICC cluster in the last 20 days of training (the Sparse 0 portfolio). 
We highlighted that the additional gain in the ICC-PO construction is mainly a consequence of the market states clustering and only partially consequence of sparsification. 
These results are extremely robust showing comparable patterns across experiments  conducted for three major capital markets, using two different solving methodologies, adopting four investment horizons and using both Student-t and normal log-likelihoods gain functions (Appendices \ref{appendixB} and \ref{appendixD}). 
The results in Appendices \ref{appendixB} and \ref{appendixD} obtained for 100 random stocks in the US, the UK and the Chinese markets show a broader variability but overall well aligned results. 
Our results also confirm the observation, by Procacci and Aste  \cite{procacci2}, that models with larger likelihood better solve the portfolio optimization problem.

As for the analysis on the 100 NASDAQ's most capitalized stocks, also for the random selection and the three markets we observe that the Normal log-likelihood, Sparse 0 is $33.8\%$ less than Full in the 5th percentile Return, whereas the Student-t log-likelihood is only $20.4\%$, which illustrates that the Normal statistically loses less money in extreme situations. 
Namely, it results that there are general advantages in using Student-t over Normal log-likelihood, yet, the latter performs a better at limiting risks. The edge in three main performance matrices depicts the Student-t's better market modelling property as suggested in the literature, especially for limited sample daily log-return. In contrast, the mere pitfall in risk measures may probably come from the  fat-tail nature of the Student-t distribution.

It is difficult to assess the efficiency of ICC-PO by direct comparison to the literature, since our focused result is the relative difference between Sparse 0 from Full. 
The most informative measurements to the general performance used widely in the field of portfolio management are Sharpe Ratio (the risk adjusted return), Jensen's Alpha (the abnormal return over the theoretical expectation), Treynor Ratio (the risk adjusted excess return from a risk-free asset) and Roy Ratio (the risk adjusted excess return from the market index) \cite{PortfolioAnalysis1}. 
Literature identifies that Sharpe Ratio at values around 1 is commonly considered as the boundary between a good and bad investment strategy, while Sharpe Ratio at values around 2 represents an excellent standard, and 3 and above are more likely to be achieved in a High Frequency Trading (HFT) strategy \cite{SR0,SR1}. 
During the 10-year we investigated the annualized Sharpe Ratio for NASDAQ-100, FTSE-250 and HS300 have been respectively equal to $1.77$, $0.42$ and $1.07$  \cite{IndexSR1,IndexSR2,IndexSR3}. 
While our results for the various portfolio construction combinations generally lie in a reasonable range around these values, we note that 
the average Sharpe Ratio of the Sparse 0 based on the 100 largest market capitalization stocks from NASDAQ is $2.425$, as well as 100 random stocks from NASDAQ is $2.132$, from FTSE is $1.682$ and from HS300 is $1.814$ greatly exceeding the index's performances. 

Apart from Sharpe Ratio for general performance assessment, risk is often a critical consideration in portfolio investment due to the risk aversion nature of investors and the quadratic utility function assumption. Two widely used risk measures are value at risk (VaR) \cite{VaR0} and probable maximum loss (PML) \cite{PML0}, which are interpreted as the minimum and the maximum loss expected in a portfolio over a time period. As a proxy combination of VaR and PML, we reported the 5th percentile Return in the random re-sampling. The observed general $66.0\%$, $49.8\%$, $21.6\%$ and $32.4\%$ reductions in loss respectively for largest-market-capitalization NASDAQ, NASDAQ, FTSE and HS300 are a highly significant result indicating likely large improvements of both VaR and PML. 

Lastly, as ICC-PO is computationally very efficient, it can be easily re-run for every allocation window making dynamic portfolio allocation easy.

\section{Conclusions}
Portfolio optimization lays at the core of quantitative investment. Automation in the dynamic allocation process is a challenging goal with a large community of academics and practitioners dedicated to this task which requires a precise and accurate modelling of the past market performance and a predictive inference of the future market state. 
However, it is never an easy task to predict the future, not to mention doing so constantly.  Explanatory as they are, only certain signals possess the forecasting ability and normally only for a limited period of time. Hence, the results of our proposed algorithm ICC-PO are worth to be mentioned. Indeed, we improve the equal weight benchmark by over 50\% in Sharpe Ratio, obtaining a statistically more robust and resilient investment performance especially in the extreme market situations with large reductions in losses. 
 
 In this paper we demonstrated that markets can be classified in different states with distinct statistical properties.  
 By using two states, classified and clustered using log-likelihood as gain function  and sparse inverse covariance estimation, we have shown that the two clustered states continue to be distinguishable in log-likelihood after the train (in-sample) period, with one having systematically larger log-likelihood than the one computed from the whole, unclustered, training sample. We have shown that the state with larger log-likelihood tends to be the one also with largest likelihood in the last period of training, indicating temporal persistence and providing a way for predictability of the best performing state in the off-sample investment period. 
 Portfolios optimized with data from the best performing state's cluster give significantly better results than portfolios constructed from the full dataset or the other state. 
 This also confirm the intuitive argument (see \cite{procacci2}) that a model with larger likelihood must perform better for portfolio optimization purposes than a model with lower likelihood. 
 These results were tested extensively across a period of ten years, across three different markets, with portfolios from two different optimizers,  with clustering from two different log-likelihoods, and both by using a selected group of most capitalized stocks as well as by random picking a stock basket. 
 
 The choice of using two market states has been dictated by simplicity. Future work will investigate the effect of the number of ICC clusters on the results. 
Our results are based on a naive selection of stocks from major indices. Hence, with a carefully designed portfolio basket, as commonly done in industrial practices, we expect further improvement of the results. Also a wider application in asset classes is a straightforward extension of the method.

\newpage

\begin{appendices}
\section{Off sample log-likelihood and performances for Student-t log-likelihood construction}\label{appendixA}

\begin{figure}[H]
        \centering
        \includegraphics[scale=0.5]{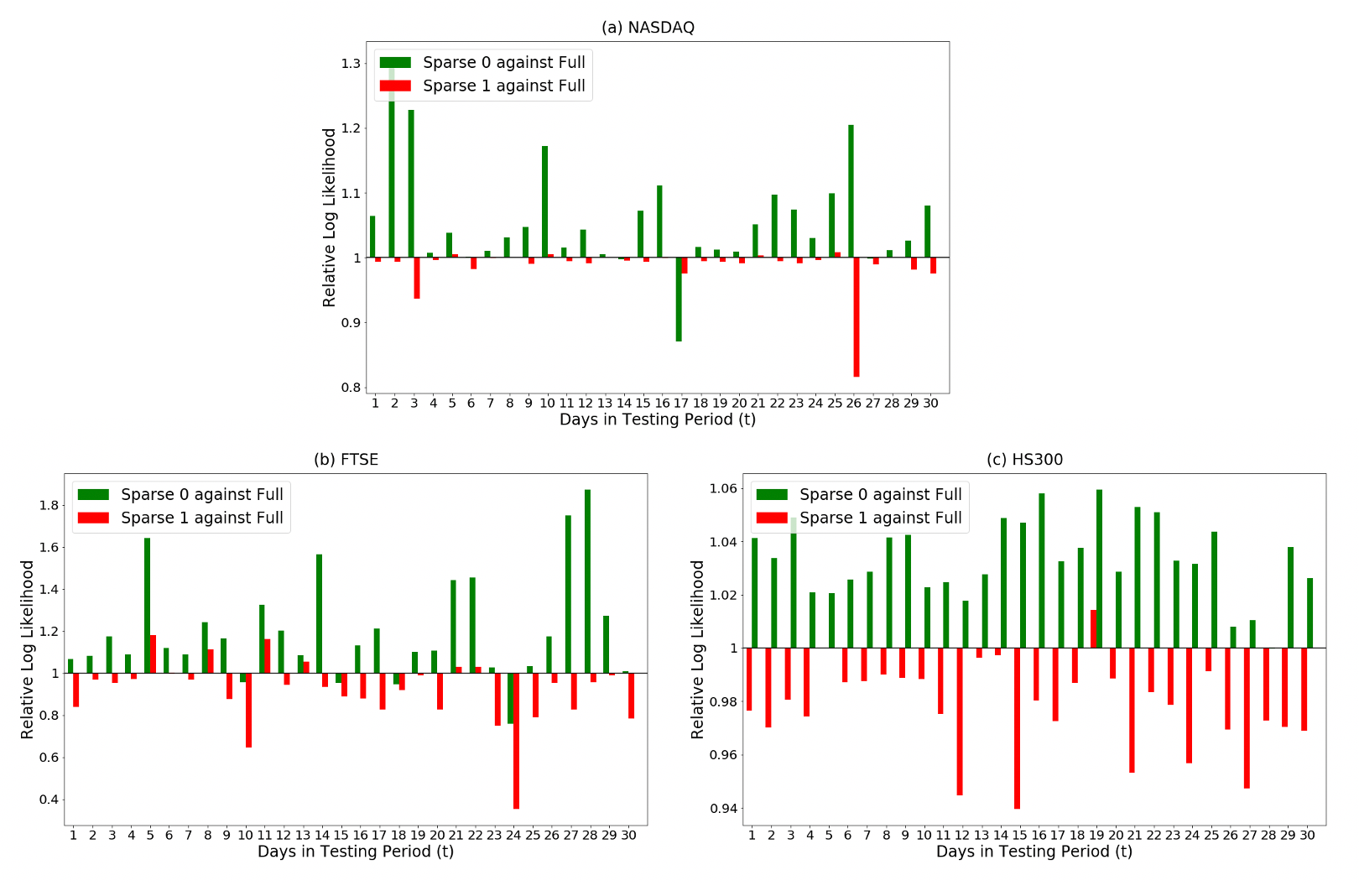}
        \caption[Likelihood plots]{
        Student-t log-likelihood for constituent stocks of a) NASDAQ, b) FTSE and c) HS300 Composite v.s. number of days in the test period after training. 
Each bar represents the average gain of the Sparse 0 (green) or 1 (red) with respect to the Full in each day. Averages are over 100 re-samplings.}
         \label{fig:ST_Likelihood_plots}
\end{figure}

In this appendix we investigate the effect of the length of the train set on the Sharpe Ratio performance and  off-sample (test set) log-likelihood using  100 randomly selected  stocks drawn from NASDAQ, FTSE and HS300. They are in the similar format as Figure \ref{fig:Selected training duration} and \ref{fig:Selected_Likelihood_plots} and demonstrate that identical patterns exist regardless underlying assets and capital markets. In Figure \ref{fig:ST_Likelihood_plots},  it is noticeable that the green bars in general sit above 0 and the red are below 0, which indicates the Sparse 0 has better off-sample log-likelihood than the Full, as illustrated in Figure \ref{fig:Selected_Likelihood_plots}.
In this appendix we compute Student-t likelihoods. 

\begin{figure}[H]
        \centering
        \includegraphics[scale=0.4]{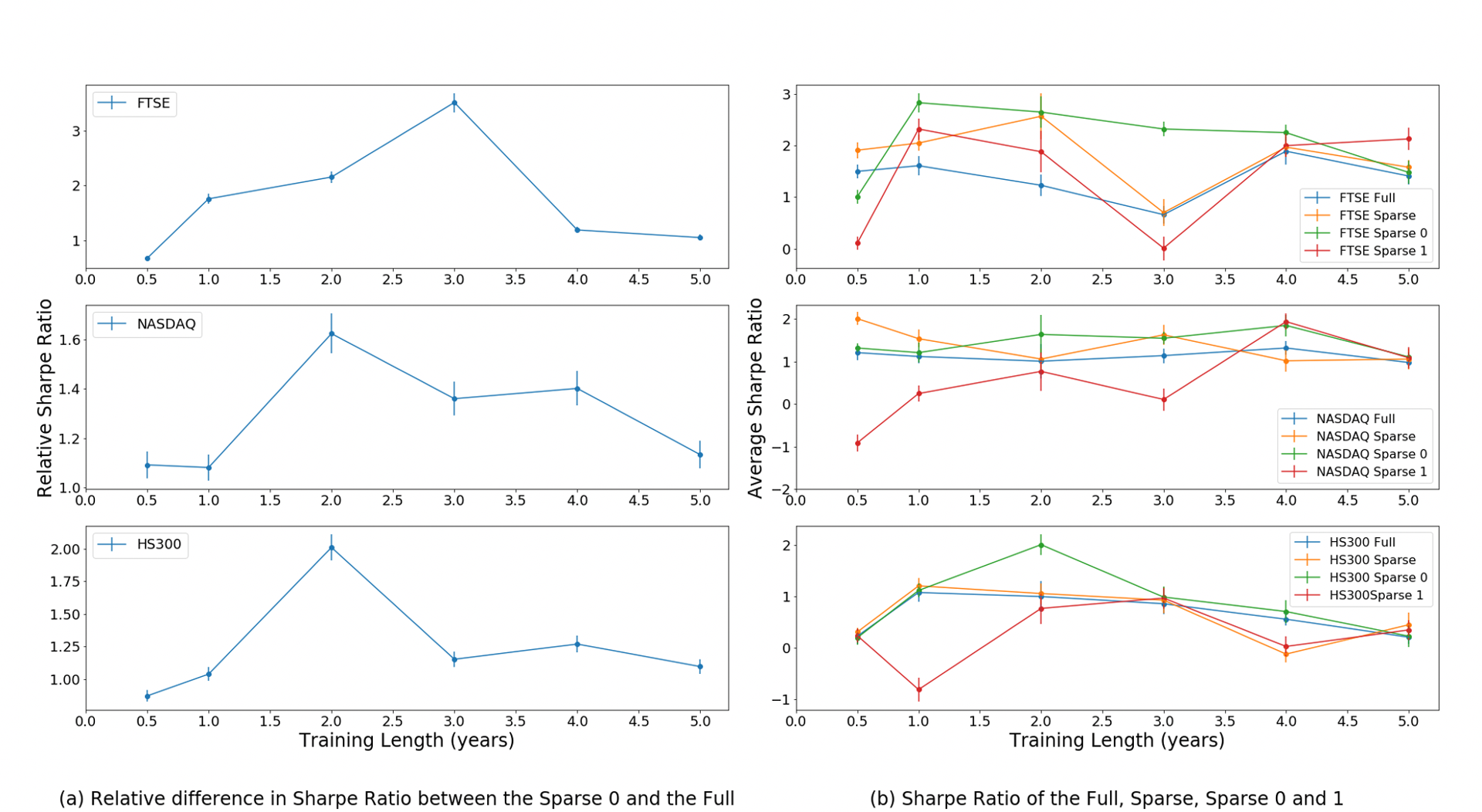}
        \caption[training duration]{Sharpe Ratio for portfolios with constituent stocks of three indices optimized using different training set durations by using Student-t log-likelihood for ICC clustering. The right subplot reports the average Sharpe Ratios ($SR$) with 1 standard deviation for states, statistics is on 100 training-testing periods chosen at random within the 10 years dataset.
        The left subplot report instead the relative Sharpe Ratios between Sparse 0 and Full, $SR_{Sparse 0}/SR_{Full}$.  
        }
         \label{fig:ST training duration}
\end{figure}
\newpage

\newpage

\section{Portfolio Performances }\label{appendixB}
In this appendix we extend the results in the main paper including   10, 20, 30 and 100-day investment horizons based on Student-t log-likelihood. 
Differently from the main text, portfolio are constructed with 100 random stocks drawn from NASDAQ, FTSE and HS300. They serve as complement and comparison for table \ref{Tab:ST_Selected30}. It is noticeable that although shorter testing period yields numerically larger Sharpe Ratio due to a possible low Volatility and a overestimation of annualized Return on small sample, the relative difference, namely, the gain between the Sparse 0 and the Full remains roughly the same. This consistency further confirms the generality of our model. Besides, the patterns in three tables are generally consistent to the findings in section \ref{portfolio_results}.

\begin{table}[H]
    \centering
\begin{tabular}{ccc|c|c|c}
\toprule
 Market & Solver &     State & Return (5, 95) percentile & Volatility (5, 95) percentile & Sharpe    (5, 95) percentile \\
  \midrule
 NASDAQ &  &  $\frac{1}{n}$ Naive  &  (-171,206) &                        (14.0,40.0) &                 (-8.7,12.5) \\
\midrule
 NASDAQ &    SLS &      Full &                     (-210,138) &                        (13.0,69.0) &                  (-7.2,9.1) \\
 NASDAQ &    SLS &    Sparse &                     (-213,122) &                        (12.0,59.0) &                  (-7.6,9.1) \\
 NASDAQ &    SLS &  Sparse 0 &                     (-157,125) &                         (9.0,52.0) &                 (-5.1,10.4) \\
 NASDAQ &    SLS &  Sparse 1 &                     (-352,153) &                        (13.0,49.0) &                  (-7.1,7.6) \\
 \midrule
 NASDAQ &    CLA &      Full &                     (-198,190) &                        (13.0,57.0) &                 (-6.6,10.1) \\
 NASDAQ &    CLA &    Sparse &                     (-172,223) &                        (13.0,51.0) &                 (-8.4,10.3) \\
 NASDAQ &    CLA &  Sparse 0 &                     (-181,185) &                        (10.0,49.0) &                 (-5.2,13.4) \\
 NASDAQ &    CLA &  Sparse 1 &                     (-198,200) &                        (13.0,66.0) &                  (-8.3,7.4) \\

 \midrule
   FTSE &     &    $\frac{1}{n}$ Naive &                     (-125,140) &                         (9.0,28.0) &                 (-9.0,17.0) \\
    \midrule
   FTSE &    SLS &      Full &                      (-91,147) &                         (7.0,27.0) &                 (-8.1,20.5) \\
   FTSE &    SLS &    Sparse &                      (-89,125) &                         (7.0,26.0) &                 (-8.8,15.5) \\
   FTSE &    SLS &  Sparse 0 &                      (-64,150) &                         (8.0,22.0) &                 (-5.9,18.8) \\
   FTSE &    SLS &  Sparse 1 &                     (-113,116) &                         (9.0,26.0) &                 (-8.9,11.6) \\
   \midrule
   
   FTSE &    CLA &      Full &                     (-147,137) &                         (8.0,22.0) &                 (-6.9,17.6) \\
   FTSE &    CLA &    Sparse &                     (-138,122) &                         (7.0,25.0) &                 (-9.0,18.0) \\
   FTSE &    CLA &  Sparse 0 &                     (-119,129) &                         (7.0,23.0) &                 (-6.4,21.5) \\
   FTSE &    CLA &  Sparse 1 &                     (-171,149) &                         (9.0,22.0) &                (-10.5,16.3) \\
   \midrule

  HS300 &     &    $\frac{1}{n}$ Naive &                     (-228,198) &                        (10.0,60.0) &                 (-7.7,10.8) \\
  \midrule
  HS300 &    SLS &      Full &                     (-250,216) &                        (12.0,42.0) &                 (-8.3,15.8) \\
  HS300 &    SLS &    Sparse &                     (-283,252) &                        (11.0,44.0) &                 (-8.1,15.3) \\
  HS300 &    SLS &  Sparse 0 &                     (-181,284) &                        (11.0,46.0) &                 (-6.0,16.0) \\
  HS300 &    SLS &  Sparse 1 &                     (-317,192) &                        (13.0,58.0) &                  (-7.9,9.9) \\
  \midrule
  
  HS300 &    CLA &      Full &                     (-250,216) &                        (12.0,42.0) &                 (-8.3,15.8) \\
  HS300 &    CLA &    Sparse &                     (-283,252) &                        (11.0,44.0) &                 (-8.1,15.3) \\
  HS300 &    CLA &  Sparse 0 &                     (-194,284) &                        (11.0,41.0) &                 (-4.9,14.2) \\
  HS300 &    CLA &  Sparse 1 &                     (-277,223) &                        (13.0,53.0) &                  (-9.2,8.6) \\

\bottomrule
  \\
  \\
    \end{tabular}
    \vspace{0.5pt}
    \caption{Portfolio performances obtained by using Student-t log-likelihood for ICC clustering.
 We report annualized  return, annualized volatility and annualized Sharpe Ratio computed on 10 days investment period after the 1 year training set.
The values are averages and 5th and 95th percentiles computed over 10-day investment horizon from obtained from 100 re-sampling of  consecutive training-investment periods chosen at random within the 10 years dataset. 
 The underlying assets are constituent stocks of NASDAQ, FTSE and HS300. Highlight in bold are return, volatility and Sharpe Ratio indicating the optimal state in each market solver combination, while highlights in 5th return and 95th volatility showcase the extreme behaviours (excluding the state Market). The state $1/n$ Naive is the equally weighted un-optimised portfolio and it is reported as benchmark. }
    \label{Tab:StudentT10}
\end{table}

\begin{table}[H]
    \centering
\begin{tabular}{ccc|c |c|c}
\toprule
 Market & Solver &     State & Return (5, 95) percentile & Volatility (5, 95) percentile & Sharpe    (5, 95) percentile \\
\midrule
 NASDAQ &     &    $\frac{1}{n}$ Naive &                     (-112,180) &                        (14.0,84.0) &                  (-4.7,5.9) \\
 \midrule
 NASDAQ &    SLS &      Full &                     (-126,131) &                        (14.0,85.0) &                  (-4.4,6.5) \\
 NASDAQ &    SLS &    Sparse &                     (-133,127) &                        (13.0,53.0) &                  (-5.1,6.5) \\
 NASDAQ &    SLS &  Sparse 0 &                     (-124,120) &                        (12.0,83.0) &                  (-4.4,7.5) \\
 NASDAQ &    SLS &  Sparse 1 &                      (-198,78) &                        (15.0,50.0) &                  (-6.1,4.5) \\
 \midrule

 NASDAQ &    CLA &      Full &                     (-147,127) &                        (13.0,84.0) &                  (-4.7,5.4) \\
 NASDAQ &    CLA &    Sparse &                     (-149,135) &                        (13.0,53.0) &                  (-4.6,6.4) \\
 NASDAQ &    CLA &  Sparse 0 &                     (-101,144) &                        (12.0,46.0) &                  (-2.8,7.9) \\
 NASDAQ &    CLA &  Sparse 1 &                      (-187,88) &                        (15.0,62.0) &                  (-5.4,4.1) \\
   
 \midrule
   FTSE &     &    $\frac{1}{n}$ Naive &                      (-77,111) &                        (11.0,26.0) &                  (-4.7,8.1) \\
   \midrule
   FTSE &    SLS &      Full &                       (-58,94) &                         (9.0,26.0) &                 (-4.5,11.4) \\
   FTSE &    SLS &    Sparse &                       (-59,95) &                        (10.0,25.0) &                 (-4.9,10.6) \\
   FTSE &    SLS &  Sparse 0 &                       (-39,96) &                         (9.0,17.0) &                 (-3.4,11.4) \\
   FTSE &    SLS &  Sparse 1 &                       (-72,75) &                         (9.0,22.0) &                  (-5.4,7.5) \\
   \midrule

   FTSE &    CLA &      Full &                       (-82,84) &                        (10.0,25.0) &                 (-5.6,10.3) \\
   FTSE &    CLA &    Sparse &                       (-62,81) &                        (10.0,20.0) &                 (-6.2,10.1) \\
   FTSE &    CLA &  Sparse 0 &                       (-59,79) &                         (9.0,22.0) &                 (-4.3,11.7) \\
   FTSE &    CLA &  Sparse 1 &                      (-100,80) &                        (10.0,23.0) &                  (-6.0,9.0) \\
   \midrule

  HS300 &     &    $\frac{1}{n}$ Naive &                     (-102,236) &                        (11.0,44.0) &                  (-4.4,9.3) \\
  \midrule
  HS300 &    SLS &      Full &                     (-133,246) &                        (15.0,42.0) &                 (-5.1,10.7) \\
  HS300 &    SLS &    Sparse &                     (-125,234) &                        (14.0,42.0) &                 (-5.0,10.3) \\
  HS300 &    SLS &  Sparse 0 &                     (-101,218) &                        (13.0,40.0) &                 (-2.8,11.3) \\
  HS300 &    SLS &  Sparse 1 &                     (-142,202) &                        (13.0,46.0) &                  (-5.2,7.9) \\
  \midrule

  HS300 &    CLA &      Full &                     (-133,246) &                        (15.0,42.0) &                 (-5.1,10.7) \\
  HS300 &    CLA &    Sparse &                     (-125,234) &                        (14.0,42.0) &                 (-5.0,10.3) \\
  HS300 &    CLA &  Sparse 0 &                      (-62,247) &                        (12.0,41.0) &                 (-2.8,10.2) \\
  HS300 &    CLA &  Sparse 1 &                     (-131,187) &                        (14.0,48.0) &                  (-5.0,8.2) \\

\bottomrule
  \\
  \\
    \end{tabular}
    \vspace{0.5pt}
    \caption{Portfolio performances obtained by using Student-t log-likelihood for ICC clustering.
 We report annualized  return, annualized volatility and annualized Sharpe Ratio computed on 20 days investment period after the 1 year training set.
The values are averages and 5th and 95th percentiles computed over 20-day investment horizon from obtained from 100  re-sampling of  consecutive training-investment periods chosen at random within the 10 years dataset. 
 The underlying assets are constituent stocks of NASDAQ, FTSE and HS300. Highlight in bold are return, volatility and Sharpe Ratio indicating the optimal state in each market solver combination, while highlights in 5th return and 95th volatility showcase the extreme behaviours (excluding the state Market). The state $1/n$ Naive is the equally weighted un-optimised portfolio and it is reported as benchmark. }

    \label{Tab:StudentT20}
\end{table}

\begin{table}[H]
    \centering
\begin{tabular}{ccc|c |c|c}
\toprule
 Market & Solver &     State & Return (5, 95) percentile & Volatility (5, 95) percentile & Sharpe    (5, 95) percentile \\
\midrule
 NASDAQ &     &    $\frac{1}{n}$ Naive &                      (-110,62) &                        (13.0,35.0) &                  (-4.5,5.7) \\
 \midrule

 NASDAQ &    SLS &    Sparse &                     (-133,115) &                        (14.0,73.0) &                  (-4.7,4.7) \\
 NASDAQ &    SLS &  Sparse 0 &                      (-99,112) &                        (14.0,71.0) &                  (-2.9,6.4) \\
 NASDAQ &    SLS &  Sparse 1 &                      (-128,72) &                        (14.0,63.0) &                  (-4.4,4.4) \\
 \midrule

 NASDAQ &    CLA &      Full &                      (-87,121) &                        (16.0,69.0) &                  (-3.0,5.8) \\
 NASDAQ &    CLA &    Sparse &                      (-86,130) &                        (15.0,72.0) &                  (-2.6,6.2) \\
 NASDAQ &    CLA &  Sparse 0 &                      (-40,134) &                        (14.0,74.0) &                  (-2.5,6.5) \\
 NASDAQ &    CLA &  Sparse 1 &                      (-101,86) &                        (15.0,73.0) &                  (-3.3,4.0) \\
 \midrule

   FTSE &     &    $\frac{1}{n}$ Naive &                       (-69,90) &                        (11.0,28.0) &                  (-3.0,6.6) \\
   \midrule
   FTSE &    SLS &      Full &                       (-63,80) &                        (10.0,26.0) &                  (-4.7,8.0) \\
   FTSE &    SLS &    Sparse &                       (-56,73) &                        (10.0,22.0) &                  (-4.8,8.3) \\
   FTSE &    SLS &  Sparse 0 &                       (-52,87) &                         (9.0,20.0) &                  (-3.5,7.3) \\
   FTSE &    SLS &  Sparse 1 &                       (-68,62) &                        (11.0,22.0) &                  (-4.5,6.3) \\
   \midrule

   FTSE &    CLA &      Full &                       (-56,79) &                        (10.0,24.0) &                  (-4.8,8.0) \\
   FTSE &    CLA &    Sparse &                       (-53,73) &                        (10.0,20.0) &                  (-4.6,9.1) \\
   FTSE &    CLA &  Sparse 0 &                       (-47,72) &                         (9.0,20.0) &                  (-4.2,9.0) \\
   FTSE &    CLA &  Sparse 1 &                       (-81,64) &                        (11.0,24.0) &                  (-5.8,7.1) \\
   \midrule

  HS300 &     &    $\frac{1}{n}$ Naive &                      (-90,172) &                        (11.0,38.0) &                  (-3.3,7.1) \\
  \midrule
  HS300 &    SLS &      Full &                     (-127,173) &                        (16.0,40.0) &                  (-4.0,6.3) \\
  HS300 &    SLS &    Sparse &                      (-98,160) &                        (15.0,36.0) &                  (-4.1,7.2) \\
  HS300 &    SLS &  Sparse 0 &                      (-65,172) &                        (13.0,43.0) &                  (-2.9,7.3) \\
  HS300 &    SLS &  Sparse 1 &                     (-118,142) &                        (15.0,45.0) &                  (-3.8,5.7) \\
  \midrule

  HS300 &    CLA &      Full &                     (-127,173) &                        (16.0,40.0) &                  (-4.0,6.2) \\
  HS300 &    CLA &    Sparse &                      (-98,160) &                        (15.0,36.0) &                  (-4.1,7.2) \\
  HS300 &    CLA &  Sparse 0 &                      (-64,173) &                        (13.0,38.0) &                  (-2.7,7.4) \\
  HS300 &    CLA &  Sparse 1 &                      (-98,142) &                        (15.0,36.0) &                  (-4.2,5.4) \\

\bottomrule
  \\
  \\
    \end{tabular}
    \vspace{0.5pt}
    \caption{Portfolio performances obtained by using Student-t log-likelihood for ICC clustering.
 We report annualized  return, annualized volatility and annualized Sharpe Ratio computed on 30 days investment period after the 1 year training set.
The values are averages and 5th and 95th percentiles computed over 30-day investment horizon from obtained from 100  re-sampling of  consecutive training-investment periods chosen at random within the 10 years dataset. 
 The underlying assets are constituent stocks of NASDAQ, FTSE and HS300. Highlight in bold are return, volatility and Sharpe Ratio indicating the optimal state in each market solver combination, while highlights in 5th return and 95th volatility showcase the extreme behaviours (excluding the state Market). The state $1/n$ Naive is the equally weighted un-optimised portfolio and it is reported as benchmark. 
}
    \label{Tab:StudentT30}
\end{table}

\begin{table}[H]
    \centering
\begin{tabular}{ccc|c |c|c}
\toprule
 Market & Solver &     State & Return (5, 95) percentile & Volatility (5, 95) percentile & Sharpe    (5, 95) percentile \\
\midrule
 NASDAQ &     &    $\frac{1}{n}$ Naive&                     (-167,172) &                        (11.0,48.0) &                  (-7.6,8.4) \\
 \midrule
 NASDAQ &    SLS &      Full &                     (-210,138) &                        (13.0,69.0) &                  (-7.2,9.1) \\
 NASDAQ &    SLS &    Sparse &                     (-213,122) &                        (12.0,59.0) &                  (-7.6,9.1) \\
 NASDAQ &    SLS &  Sparse 0 &                     (-189,160) &                        (11.0,41.0) &                  (-5.4,8.4) \\
 NASDAQ &    SLS &  Sparse 1 &                     (-290,114) &                        (13.0,66.0) &                  (-8.5,5.6) \\
 \midrule

 NASDAQ &     &      Full &                     (-198,190) &                        (13.0,57.0) &                 (-6.6,10.1) \\
 NASDAQ &    CLA &    Sparse &                     (-172,223) &                        (13.0,51.0) &                 (-8.4,10.3) \\
 NASDAQ &    CLA &  Sparse 0 &                     (-165,187) &                        (11.0,46.0) &                 (-6.1,14.2) \\
 NASDAQ &    CLA &  Sparse 1 &                     (-257,166) &                        (13.0,62.0) &                  (-6.8,6.6) \\
 \midrule

   FTSE &      &    $\frac{1}{n}$ Naive &                     (-186,140) &                         (8.0,22.0) &                (-10.3,23.0) \\
    \midrule
   FTSE &    SLS &      Full &                      (-91,147) &                         (7.0,27.0) &                 (-8.1,20.5) \\
   FTSE &    SLS &    Sparse &                      (-89,125) &                         (7.0,26.0) &                 (-8.8,15.5) \\
   FTSE &    SLS &  Sparse 0 &                      (-75,161) &                         (7.0,21.0) &                 (-7.4,29.2) \\
   FTSE &    SLS &  Sparse 1 &                     (-110,123) &                         (9.0,27.0) &                 (-9.5,13.6) \\
   \midrule

   FTSE &    CLA &      Full &                     (-147,137) &                         (8.0,22.0) &                 (-6.9,17.6) \\
   FTSE &    CLA &    Sparse &                     (-138,122) &                         (7.0,25.0) &                 (-9.0,18.0) \\
   FTSE &    CLA &  Sparse 0 &                      (-80,145) &                         (8.0,23.0) &                 (-5.7,24.2) \\
   FTSE &    CLA &  Sparse 1 &                     (-194,138) &                         (9.0,25.0) &                (-12.0,16.3) \\
   \midrule

  HS300 &     &    $\frac{1}{n}$ Naive &                     (-228,198) &                        (10.0,60.0) &                 (-7.7,10.7) \\
  \midrule
  HS300 &    SLS &      Full &                     (-250,216) &                        (12.0,42.0) &                 (-8.3,15.8) \\
  HS300 &    SLS &    Sparse &                     (-283,252) &                        (11.0,44.0) &                 (-8.1,15.3) \\
  HS300 &    SLS &  Sparse 0 &                     (-237,249) &                        (11.0,50.0) &                 (-5.6,16.5) \\
  HS300 &    SLS &  Sparse 1 &                     (-237,193) &                        (14.0,45.0) &                  (-8.4,9.3) \\
  \midrule

  HS300 &    CLA &      Full &                     (-250,216) &                        (12.0,42.0) &                 (-8.3,15.8) \\
  HS300 &    CLA &    Sparse &                     (-283,252) &                        (11.0,44.0) &                 (-8.1,15.3) \\
  HS300 &    CLA &  Sparse 0 &                     (-186,298) &                        (10.0,44.0) &                 (-7.6,13.9) \\
  HS300 &    CLA &  Sparse 1 &                     (-302,186) &                        (13.0,56.0) &                  (-7.5,9.5) \\

\bottomrule
  \\
  \\
    \end{tabular}
    \vspace{0.5pt}
    \caption{Portfolio performances obtained by using Student-t log-likelihood for ICC clustering.
 We report annualized  return, annualized volatility and annualized Sharpe Ratio computed on 100 days investment period after the 1 year training set.
The values are averages and 5th and 95th percentiles computed over 100-day investment horizon from obtained from 100  re-sampling of  consecutive training-investment periods chosen at random within the 10 years dataset. 
 The underlying assets are constituent stocks of NASDAQ, FTSE and HS300. Highlight in bold are return, volatility and Sharpe Ratio indicating the optimal state in each market solver combination, while highlights in 5th return and 95th volatility showcase the extreme behaviours (excluding the state Market). The state $1/n$ Naive is the equally weighted un-optimised portfolio and it is reported as benchmark. 
}
    \label{Tab:StudentT100}
\end{table}

\newpage

\section{Normal Log-likelihood: training duration and Off-sample Log-likelihood}\label{appendixC}

This Appendix C section includes Sharpe ratio against training duration plots and off-sample Normal log-likelihood plots of 100 random stocks drawn from NASDAQ, FTSE and HS300. They are in the similar format as Figure \ref{fig:Selected training duration} and \ref{fig:Selected_Likelihood_plots} and demonstrate that identical patterns exist regardless underlying assets and capital markets. In Figure \ref{fig:ST_Likelihood_plots},  it is noticeable that the green bars in general sit above 0 and the red are below 0, which indicates the Sparse 0 has better off-sample log-likelihood than the Full, as illustrated in Figure \ref{fig:Selected_Likelihood_plots}.

\begin{figure}[H]
        \centering
        \includegraphics[scale=0.4]{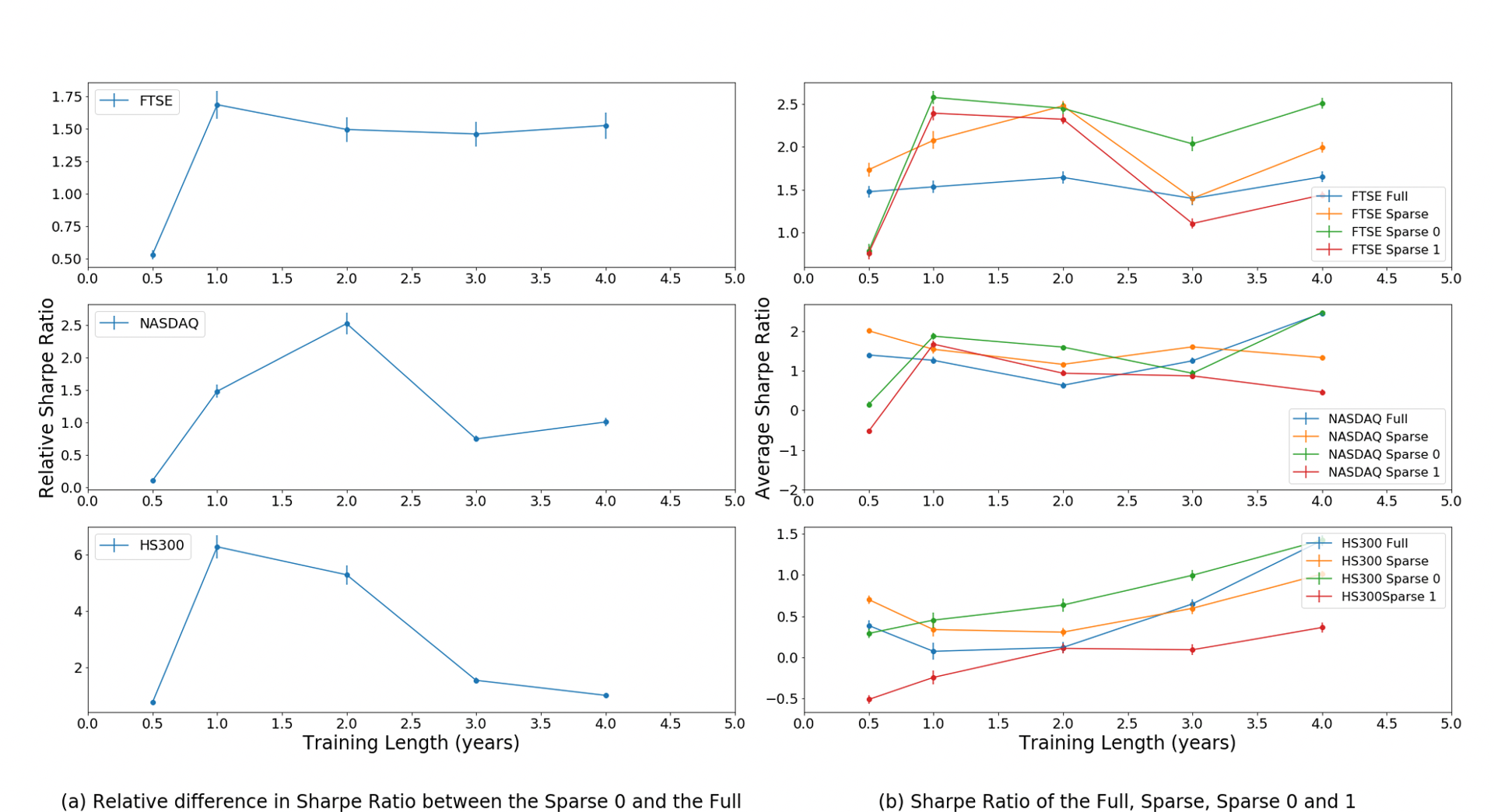}
        
        \vspace{0.5pt}
        \caption[training duration]{Sharpe Ratio for portfolios with constituent stocks of three indices optimized using different training set durations by using Normal log-likelihood for ICC clustering. The right subplot reports the average Sharpe Ratios ($SR$) with 1 standard deviation for states, statistics is on 100 training-testing periods chosen at random within the 10 years dataset.
        The left subplot report instead the relative Sharpe Ratios between Sparse 0 and Full, $SR_{Sparse 0}/SR_{Full}$.}
         \label{fig:Normal training duration}
\end{figure}
\newpage
\begin{figure}[H]
        \centering
        \includegraphics[scale=0.5]{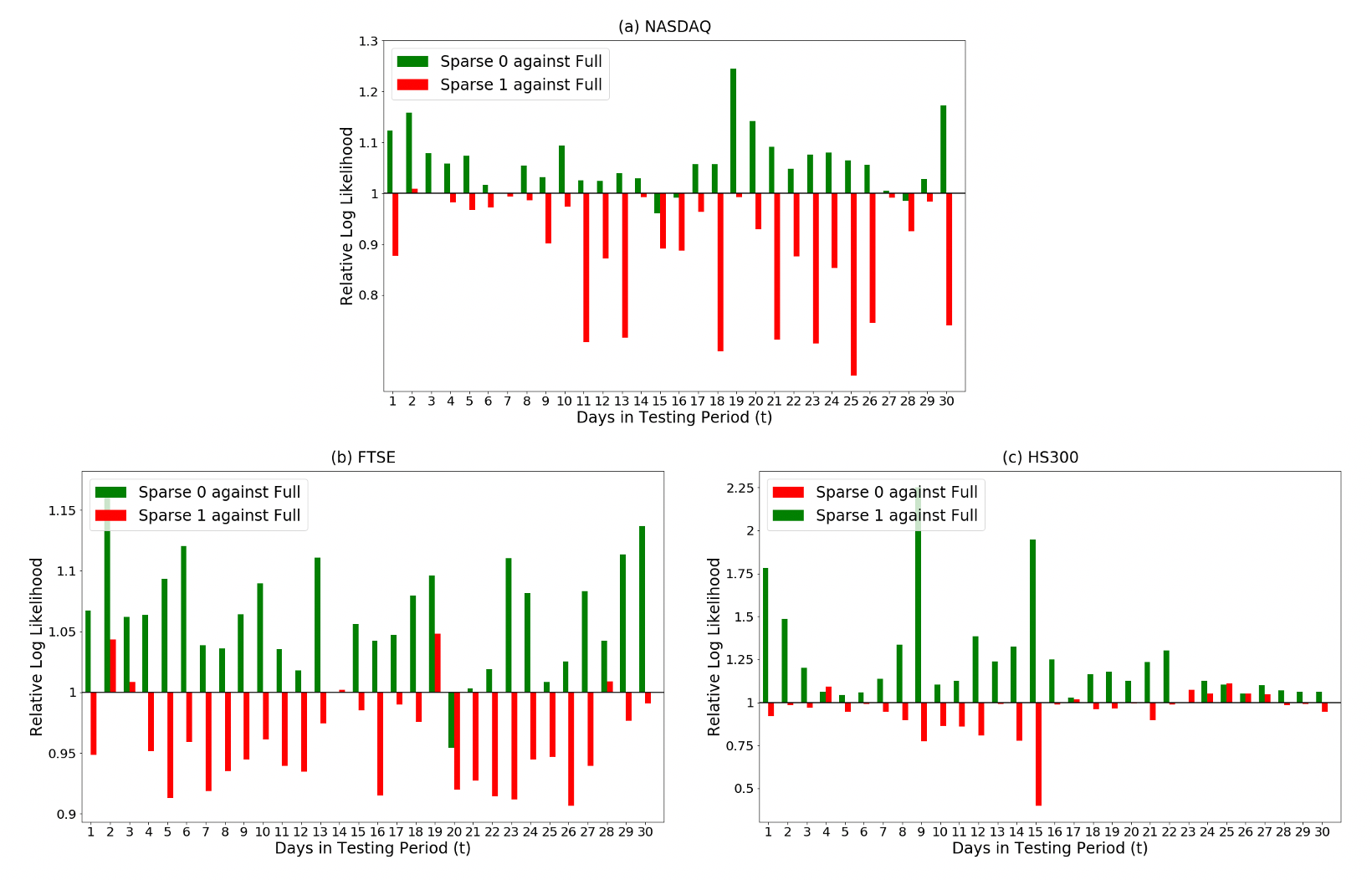}
        
        \vspace{0.5pt}
        \caption[Likelihood plots]{
        Normal log-likelihood for constituent stocks of a) NASDAQ, b) FTSE and c) HS300 Composite v.s. number of days in the test period after training. 
Each bar represents the average gain of the Sparse 0 (green) or 1 (red) with respect to the Full in each day. Averages are over 100 re-samplings.}
         \label{fig:Normal_Likelihood_plots}
\end{figure}

\newpage

\section{Off sample log-likelihood and performances for Normal log-likelihood construction}\label{appendixD}

In this appendix we perform the same kind of investigations as in the previous appendix but ICC is computed using Normal log-likelihood.  We notice similar patterns but the Student-t log-likelihood result are more significant. However, the Normal log-likelihood performs better in risk matrices.

\begin{table}[H]
    \centering
\begin{tabular}{ccc|c |c|c}
\toprule
 Market & Solver &     State & Return (5, 95) percentile & Volatility (5, 95) percentile & Sharpe    (5, 95) percentile \\
\midrule
 NASDAQ &     &    $\frac{1}{n}$ Naive &                     (-171,206) &                        (14.0,40.0) &                 (-8.7,12.5) \\
 \midrule
 NASDAQ &    SLS &      Full &                     (-192,190) &                        (13.0,57.0) &                 (-5.4,10.1) \\
 NASDAQ &    SLS &    Sparse &                     (-160,223) &                        (13.0,51.0) &                 (-7.4,10.3) \\
 NASDAQ &    SLS &  Sparse 0 &                     (-144,174) &                        (11.0,49.0) &                 (-5.0,12.6) \\
 NASDAQ &    SLS &  Sparse 1 &                     (-181,218) &                        (14.0,59.0) &                  (-8.7,7.8) \\
 \midrule

 NASDAQ &    CLA &      Full &                     (-192,198) &                        (13.0,57.0) &                 (-5.4,10.1) \\
 NASDAQ &    CLA &    Sparse &                     (-160,223) &                        (13.0,51.0) &                 (-7.4,10.3) \\
 NASDAQ &    CLA &  Sparse 0 &                     (-169,171) &                        (12.0,36.0) &                 (-5.0,14.2) \\
 NASDAQ &    CLA &  Sparse 1 &                     (-256,144) &                        (14.0,67.0) &                  (-6.5,6.5) \\
 \midrule

   FTSE &     &    $\frac{1}{n}$ Naive &                     (-161,117) &                         (7.0,34.0) &                 (-9.2,15.4) \\
   \midrule
   FTSE &    SLS &      Full &                     (-163,116) &                         (8.0,28.0) &                 (-8.7,14.1) \\
   FTSE &    SLS &    Sparse &                     (-148,108) &                         (8.0,33.0) &                 (-8.9,14.2) \\
   FTSE &    SLS &  Sparse 0 &                     (-111,138) &                         (7.0,22.0) &                 (-7.1,18.8) \\
   FTSE &    SLS &  Sparse 1 &                     (-199,118) &                         (9.0,40.0) &                (-12.1,13.0) \\
   \midrule

   FTSE &    CLA &      Full &                     (-163,116) &                         (8.0,28.0) &                 (-8.7,14.1) \\
   FTSE &    CLA &    Sparse &                     (-148,108) &                         (8.0,33.0) &                 (-8.9,14.2) \\
   FTSE &    CLA &  Sparse 0 &                     (-111,146) &                         (7.0,21.0) &                 (-8.3,18.0) \\
   FTSE &    CLA &  Sparse 1 &                     (-176,123) &                         (8.0,36.0) &                (-11.0,13.7) \\
   \midrule

  HS300 &     &    $\frac{1}{n}$ Naive &                     (-228,198) &                        (10.0,60.0) &                 (-7.7,10.8) \\
  \midrule
  HS300 &    SLS &      Full &                     (-250,216) &                        (12.0,42.0) &                 (-8.3,15.8) \\
  HS300 &    SLS &    Sparse &                     (-283,252) &                        (11.0,44.0) &                 (-8.1,15.3) \\
  HS300 &    SLS &  Sparse 0 &                     (-165,276) &                        (11.0,42.0) &                 (-5.6,15.2) \\
  HS300 &    SLS &  Sparse 1 &                     (-289,176) &                        (13.0,44.0) &                  (-8.3,7.9) \\
  \midrule

  HS300 &    CLA &      Full &                     (-250,216) &                        (12.0,42.0) &                 (-8.3,15.8) \\
  HS300 &    CLA &    Sparse &                     (-283,252) &                        (11.0,44.0) &                 (-8.1,15.3) \\
  HS300 &    CLA &  Sparse 0 &                     (-131,250) &                        (11.0,45.0) &                 (-5.3,18.8) \\
  HS300 &    CLA &  Sparse 1 &                     (-331,218) &                        (12.0,58.0) &                  (-8.9,9.9) \\

\bottomrule
  \\
  \\
    \end{tabular}
    \vspace{0.5pt}
    \caption{Portfolio performances obtained by using Normal log-likelihood for ICC clustering.
 We report annualized  return, annualized volatility and annualized Sharpe Ratio computed on 10 days investment period after the 1 year training set.
The values are averages and 5th and 95th percentiles computed over 10-day investment horizon from obtained from 100  re-sampling of  consecutive training-investment periods chosen at random within the 10 years dataset. 
 The underlying assets are constituent stocks of NASDAQ, FTSE and HS300. Highlight in bold are return, volatility and Sharpe Ratio indicating the optimal state in each market solver combination, while highlights in 5th return and 95th volatility showcase the extreme behaviours (excluding the state Market). The state $1/n$ Naive is the equally weighted un-optimised portfolio and it is reported as benchmark. }
    \label{Tab:average performance in 10}
\end{table}

\begin{table}[H]
    \centering
\begin{tabular}{ccc|c |c|c}
\toprule
 Market & Solver &     State & Return (5, 95) percentile & Volatility (5, 95) percentile & Sharpe    (5, 95) percentile \\
\midrule
 NASDAQ &     &    $\frac{1}{n}$ Naive &                     (-129,147) &                        (14.0,36.0) &                  (-4.9,7.8) \\
 \midrule
 NASDAQ &    SLS &      Full &                     (-147,124) &                        (13.0,84.0) &                  (-4.7,5.4) \\
 NASDAQ &    SLS &    Sparse &                     (-149,135) &                        (13.0,53.0) &                  (-4.6,6.4) \\
 NASDAQ &    SLS &  Sparse 0 &                      (-98,133) &                        (12.0,64.0) &                  (-3.7,8.4) \\
 NASDAQ &    SLS &  Sparse 1 &                     (-147,101) &                        (14.0,71.0) &                  (-5.6,5.3) \\
 \midrule

 NASDAQ &    CLA &      Full &                     (-147,127) &                        (13.0,84.0) &                  (-4.7,5.4) \\
 NASDAQ &    CLA &    Sparse &                     (-149,135) &                        (13.0,53.0) &                  (-4.6,6.4) \\
 NASDAQ &    CLA &  Sparse 0 &                      (-95,127) &                        (11.0,51.0) &                  (-3.1,7.8) \\
 NASDAQ &    CLA &  Sparse 1 &                     (-149,111) &                        (14.0,69.0) &                  (-4.8,5.5) \\
 \midrule

   FTSE &     &    $\frac{1}{n}$ Naive &                      (-83,104) &                        (10.0,32.0) &                  (-6.9,9.1) \\
   \midrule
   FTSE &    SLS &      Full &                      (-79,100) &                         (9.0,30.0) &                  (-5.5,9.5) \\
   FTSE &    SLS &    Sparse &                       (-63,84) &                         (9.0,27.0) &                  (-5.6,9.6) \\
   FTSE &    SLS &  Sparse 0 &                       (-49,82) &                         (9.0,27.0) &                 (-5.7,11.7) \\
   FTSE &    SLS &  Sparse 1 &                       (-92,94) &                        (11.0,34.0) &                  (-5.6,8.0) \\
   \midrule

   FTSE &    CLA &      Full &                      (-79,100) &                         (9.0,30.0) &                  (-5.5,9.5) \\
   FTSE &    CLA &    Sparse &                       (-63,84) &                         (9.0,27.0) &                  (-5.6,9.6) \\
   FTSE &    CLA &  Sparse 0 &                       (-68,82) &                         (9.0,23.0) &                 (-5.2,10.7) \\
   FTSE &    CLA &  Sparse 1 &                     (-110,101) &                         (9.0,29.0) &                  (-7.3,8.3) \\
   \midrule

  HS300 &     &    $\frac{1}{n}$ Naive &                     (-102,236) &                        (11.0,44.0) &                  (-4.4,9.3) \\
  \midrule
  HS300 &    SLS &      Full &                     (-133,246) &                        (15.0,42.0) &                 (-5.1,10.7) \\
  HS300 &    SLS &    Sparse &                     (-125,234) &                        (14.0,42.0) &                 (-5.0,10.3) \\
  HS300 &    SLS &  Sparse 0 &                      (-92,231) &                        (13.0,45.0) &                 (-3.8,11.4) \\
  HS300 &    SLS &  Sparse 1 &                     (-143,204) &                        (14.0,43.0) &                  (-5.6,7.4) \\
  \midrule

  HS300 &    CLA &      Full &                     (-133,246) &                        (15.0,42.0) &                 (-5.1,10.7) \\
  HS300 &    CLA &    Sparse &                     (-125,234) &                        (14.0,42.0) &                 (-5.0,10.3) \\
  HS300 &    CLA &  Sparse 0 &                      (-94,223) &                        (12.0,46.0) &                 (-3.0,10.0) \\
  HS300 &    CLA &  Sparse 1 &                     (-146,209) &                        (12.0,39.0) &                  (-5.5,8.0) \\

\bottomrule
  \\
  \\
    \end{tabular}
    \vspace{0.5pt}
    \caption{Portfolio performances obtained by using Normal log-likelihood for ICC clustering.
 We report annualized  return, annualized volatility and annualized Sharpe Ratio computed on 20 days investment period after the 1 year training set.
The values are averages and 5th and 95th percentiles computed over 20-day investment horizon from obtained from 100  re-sampling of  consecutive training-investment periods chosen at random within the 10 years dataset. 
 The underlying assets are constituent stocks of NASDAQ, FTSE and HS300. Highlight in bold are return, volatility and Sharpe Ratio indicating the optimal state in each market solver combination, while highlights in 5th return and 95th volatility showcase the extreme behaviours (excluding the state Market). The state $1/n$ Naive is the equally weighted un-optimised portfolio and it is reported as benchmark. }
    \label{Tab:average performance in 20}
\end{table}

\begin{table}[H]
    \centering
\begin{tabular}{ccc|c |c|c}
\toprule
 Market & Solver &     State & Return (5, 95) percentile & Volatility (5, 95) percentile & Sharpe    (5, 95) percentile \\
\midrule
 NASDAQ &    &   $\frac{1}{n}$ Naive &                     (-112,137) &                        (15.0,41.0) &                  (-3.3,7.0) \\
 \midrule
 NASDAQ &    SLS &      Full &                     (-105,113) &                        (16.0,71.0) &                  (-3.0,5.0) \\
 NASDAQ &    SLS &    Sparse &                     (-135,120) &                        (15.0,73.0) &                  (-3.2,5.5) \\
 NASDAQ &    SLS &  Sparse 0 &                      (-52,116) &                        (12.0,78.0) &                  (-2.6,5.6) \\
 NASDAQ &    SLS &  Sparse 1 &                      (-169,86) &                        (16.0,74.0) &                  (-3.9,3.3) \\
 \midrule

 NASDAQ &    CLA &      Full &                     (-105,113) &                        (16.0,71.0) &                  (-3.0,5.0) \\
 NASDAQ &    CLA &    Sparse &                     (-135,120) &                        (15.0,73.0) &                  (-3.2,5.5) \\
 NASDAQ &    CLA &  Sparse 0 &                      (-61,116) &                        (14.0,71.0) &                  (-2.5,6.2) \\
 NASDAQ &    CLA &  Sparse 1 &                      (-146,78) &                        (15.0,79.0) &                  (-4.1,3.7) \\
 \midrule

   FTSE &    &    $\frac{1}{n}$ Naive&                       (-46,68) &                        (11.0,31.0) &                  (-3.0,5.7) \\
   \midrule
   FTSE &    SLS &      Full &                       (-45,75) &                        (11.0,26.0) &                  (-2.9,6.8) \\
   FTSE &    SLS &    Sparse &                       (-48,69) &                        (11.0,24.0) &                  (-3.3,7.7) \\
   FTSE &    SLS &  Sparse 0 &                       (-32,68) &                        (11.0,21.0) &                  (-2.6,8.2) \\
   FTSE &    SLS &  Sparse 1 &                       (-69,67) &                        (11.0,29.0) &                  (-3.8,6.6) \\
   \midrule
 
   FTSE &    CLA &      Full &                       (-45,75) &                        (11.0,26.0) &                  (-3.0,6.8) \\
   FTSE &    CLA &    Sparse &                       (-48,69) &                        (11.0,24.0) &                  (-3.3,7.7) \\
   FTSE &    CLA &  Sparse 0 &                       (-43,67) &                        (11.0,22.0) &                  (-3.0,7.6) \\
   FTSE &    CLA &  Sparse 1 &                       (-56,58) &                        (12.0,29.0) &                  (-3.2,5.3) \\
   \midrule

  HS300 &   &   $\frac{1}{n}$ Naive &                      (-91,165) &                        (11.0,51.0) &                  (-3.3,6.0) \\
  \midrule
  HS300 &    SLS &      Full &                     (-127,168) &                        (16.0,42.0) &                  (-4.0,6.3) \\
  HS300 &    SLS &    Sparse &                      (-94,163) &                        (15.0,38.0) &                  (-4.1,7.1) \\
  HS300 &    SLS &  Sparse 0 &                      (-78,182) &                        (12.0,38.0) &                  (-3.0,6.5) \\
  HS300 &    SLS &  Sparse 1 &                     (-121,135) &                        (14.0,55.0) &                  (-3.9,5.5) \\
  \midrule

  HS300 &    CLA &      Full &                     (-127,168) &                        (16.0,42.0) &                  (-4.0,6.3) \\
  HS300 &    CLA &    Sparse &                      (-94,163) &                        (15.0,38.0) &                  (-4.1,7.1) \\
  HS300 &    CLA &  Sparse 0 &                      (-54,165) &                        (12.0,51.0) &                  (-2.3,7.8) \\
  HS300 &    CLA &  Sparse 1 &                     (-110,138) &                        (14.0,43.0) &                  (-4.2,6.0) \\

\bottomrule
  \\
  \\
    \end{tabular}
    \vspace{0.5pt}
    \caption{Portfolio performances obtained by using Normal log-likelihood for ICC clustering.
 We report annualized  return, annualized volatility and annualized Sharpe Ratio computed on 30 days investment period after the 1 year training set.
The values are averages and 5th and 95th percentiles computed over 30-day investment horizon from obtained from 100  re-sampling of  consecutive training-investment periods chosen at random within the 10 years dataset. 
 The underlying assets are constituent stocks of NASDAQ, FTSE and HS300. Highlight in bold are return, volatility and Sharpe Ratio indicating the optimal state in each market solver combination, while highlights in 5th return and 95th volatility showcase the extreme behaviours (excluding the state Market). The state $1/n$ Naive is the equally weighted un-optimised portfolio and it is reported as benchmark. }
        \label{Tab:average performance in 30}

\end{table}

\begin{table}[H]
    \centering
\begin{tabular}{ccc|c |c|c}
\toprule
 Market & Solver &     State & Return (5, 95) percentile & Volatility (5, 95) percentile & Sharpe    (5, 95) percentile \\
\midrule
 NASDAQ &     &   $\frac{1}{n}$ Naive &                       (-25,42) &                        (15.0,33.0) &                  (-1.3,2.6) \\
 \midrule
 NASDAQ &    SLS &      Full &                       (-33,52) &                        (17.0,53.0) &                  (-1.4,2.8) \\
 NASDAQ &    SLS &    Sparse &                       (-24,56) &                        (16.0,36.0) &                  (-1.2,3.1) \\
 NASDAQ &    SLS &  Sparse 0 &                       (-17,38) &                        (14.0,37.0) &                  (-0.8,2.3) \\
 NASDAQ &    SLS &  Sparse 1 &                       (-41,51) &                        (17.0,51.0) &                  (-1.9,2.1) \\
 \midrule

 NASDAQ &    CLA &      Full &                       (-33,52) &                        (17.0,53.0) &                  (-1.4,2.8) \\
 NASDAQ &    CLA &    Sparse &                       (-24,56) &                        (16.0,36.0) &                  (-1.2,3.1) \\
 NASDAQ &    CLA &  Sparse 0 &                       (-24,52) &                        (15.0,36.0) &                  (-1.2,2.4) \\
 NASDAQ &    CLA &  Sparse 1 &                       (-34,34) &                        (16.0,49.0) &                  (-1.7,1.7) \\
 \midrule

   FTSE &     &   $\frac{1}{n}$ Naive &                       (-37,49) &                        (11.0,30.0) &                  (-2.3,5.3) \\
   \midrule
   FTSE &    SLS &      Full &                       (-34,55) &                        (11.0,26.0) &                  (-2.2,5.0) \\
   FTSE &    SLS &    Sparse &                       (-38,55) &                        (11.0,26.0) &                  (-2.1,5.0) \\
   FTSE &    SLS &  Sparse 0 &                       (-27,44) &                        (10.0,18.0) &                  (-1.9,6.3) \\
   FTSE &    SLS &  Sparse 1 &                       (-39,59) &                        (12.0,30.0) &                  (-2.3,5.1) \\
   \midrule

   FTSE &    CLA &      Full &                       (-34,55) &                        (11.0,26.0) &                  (-2.2,5.0) \\
   FTSE &    CLA &    Sparse &                       (-38,55) &                        (11.0,26.0) &                  (-2.1,5.0) \\
   FTSE &    CLA &  Sparse 0 &                       (-28,57) &                        (11.0,18.0) &                  (-1.9,7.2) \\
   FTSE &    CLA &  Sparse 1 &                       (-45,44) &                        (11.0,34.0) &                  (-2.6,4.5) \\
   \midrule

  HS300 &   &    $\frac{1}{n}$ Naive &                      (-47,112) &                        (15.0,51.0) &                  (-1.9,5.0) \\
  \midrule
  HS300 &    SLS &      Full &                       (-68,77) &                        (17.0,52.0) &                  (-2.0,3.8) \\
  HS300 &    SLS &    Sparse &                       (-50,96) &                        (16.0,43.0) &                  (-1.8,4.9) \\
  HS300 &    SLS &  Sparse 0 &                      (-60,109) &                        (15.0,46.0) &                  (-1.8,5.5) \\
  HS300 &    SLS &  Sparse 1 &                      (-74,114) &                        (17.0,48.0) &                  (-2.0,4.1) \\
  \midrule

  HS300 &    CLA &      Full &                       (-68,77) &                        (17.0,52.0) &                  (-2.0,3.8) \\
  HS300 &    CLA &    Sparse &                       (-50,96) &                        (16.0,43.0) &                  (-1.8,4.9) \\
  HS300 &    CLA &  Sparse 0 &                       (-54,86) &                        (15.0,51.0) &                  (-1.4,5.3) \\
  HS300 &    CLA &  Sparse 1 &                       (-71,91) &                        (17.0,46.0) &                  (-2.0,4.3) \\

\bottomrule
  \\
  \\
    \end{tabular}
    \vspace{0.5pt}
    \caption{Portfolio performances obtained by using Normal log-likelihood for ICC clustering.
 We report annualized  return, annualized volatility and annualized Sharpe Ratio computed on 100 days investment period after the 1 year training set.
The values are averages and 5th and 95th percentiles computed over 100-day investment horizon from obtained from 100  re-sampling of  consecutive training-investment periods chosen at random within the 10 years dataset. 
 The underlying assets are constituent stocks of NASDAQ, FTSE and HS300. Highlight in bold are return, volatility and Sharpe Ratio indicating the optimal state in each market solver combination, while highlights in 5th return and 95th volatility showcase the extreme behaviours (excluding the state Market). The state $1/n$ Naive is the equally weighted un-optimised portfolio and it is reported as benchmark. }
    \label{Tab:average performance in 100}
\end{table}
\newpage

\section{Portfolio Optimization} \label{appendixF}

In the original {\bf Markowitz's mean variance optimization} approach, the portfolio weights $\mathbf{W}=(w_1,...,w_n) \in \mathbb R^{1\times n}$ are chosen in order to minimize portfolio's variance $\sigma_p^2 = \mathbf{W\Sigma W}^\top$ for a given value,  of the portfolio's expected return $\boldsymbol \mu {\mathbf W}^\top = \bar{r}_p$. 
Specifically, 
\begin{equation}\label{eq:markowitz_model}
\begin{aligned}
    \mathbf W^* =  \min_{{\mathbf W}} \quad &  \mathbf{W\Sigma W}^\top\\
    \textrm{s.t } \quad & \mathbf{\mathds{1}W}^\top=1,\\
    \textrm{and }\quad & \boldsymbol \mu {\mathbf W}^\top =\bar{r}_p,
\end{aligned}
\end{equation}
The exact solution can be obtained analytically by setting to zero the derivatives with respect to $\mathbf W$, using the Lagrange multiplier technique to account for the constraints. Namely the minimum of the following Lagrangian is computed 
\begin{equation}
    L({\mathbf W},\lambda) = \mathbf{W\Sigma W}^\top + \lambda_1 \boldsymbol \mu {\mathbf W}^\top   + \lambda_2  \mathbf{\mathds{1}W}^\top ,
 \end{equation}
and the solution is
\begin{equation}\label{eq:markowitz_solution}
    \mathbf{W}^*=\mathbf{\Sigma^{-1}}(\lambda_1 \boldsymbol \mu + \lambda_2 \mathbf{\mathds{1}})^\top,
\end{equation} 
where $\lambda_1$ and $\lambda_2$ are the Lagrange multipliers. 

The {\bf sequential least square quadratic programming (SLS)} \cite{Quadratic_optimisation0,SLSQP1,SLSQP2} is considered to be one of the most efficient computational method to solve general nonlinear constrained optimization problems. Jackson et al. and Cesarone et al. demonstrate its effectiveness in finance \cite{Quadratic_optimisation1,Quadratic_optimisation2}. 
SLS solves the optimization problem iteratively with a gradient descent strategy starting with an initial setting ${\mathbf W}^0$, and updating  ${\mathbf W}^{k+1}$  from ${\mathbf W}^{k}$ by:
\begin{equation}\label{eq:SLS}
{\mathbf W}^{k+1}={\mathbf W}^{k}+\alpha^k {\mathbf d}^k
\end{equation}
where ${\mathbf d}^k$ is the search direction at the $k$-th step and $\alpha^k$ is the associated step size. In each iteration, the descent search direction, ${\mathbf d}$, is determined by the solution of a sub-problem. 
Given the loss function 
\begin{equation}
f({\mathbf W})=\mathbf{W\Sigma W}^\top
\end{equation}
that we want to minimize under a set of non-liner constraints $g_j({\mathbf W})=0$ for $j\in[1,m_e]$ and $g_j({\mathbf W})\ge 0$ for $j\in[m_e+1,m]$,
at each iteration, the problem of finding the optimal descent direction can be addressed by solving the standard quadratic programming sub-problem \cite{SLSQP3}: 
\begin{equation}\label{eq:search direction}
\begin{aligned}
\mathbf d^{k+1} = \min_{{\mathbf d}}
  \quad &{\frac{1}{2}{\mathbf d} \nabla^2 L({\mathbf W}^k,\boldsymbol \lambda) {\mathbf d}^\top+\nabla f({\mathbf W}^k){\mathbf d}^\top}\\
    \textrm{s.t }\quad & \nabla g_j({\mathbf W}^k){\mathbf d}^\top+g_j({\mathbf W}^k)=0, \;\;\; j=1,...,m_e \\
    \quad & \nabla g_j({\mathbf W}^k){\mathbf d}^\top+g_j({\mathbf W}^k)\geq 0,\;\;\;  j=m_e+1,...,m \\
\end{aligned}
\end{equation}
where $L({\mathbf W},\boldsymbol \lambda)$ is the associated Lagrangian 
\begin{equation}
    L({\mathbf W},\boldsymbol \lambda)=f({\mathbf W})-\sum_{j=1}^{m} \lambda_j g_j({\mathbf W}).
\end{equation}

A step size $\alpha =1$ is optimal near a local optimum, but when far from the optimum, the step size will need to be modified to guarantee a global convergence. Han \cite{SLSQP4}, Powell \cite{SLSQP6}, Schittkowski \cite{SLSQP7} and Rockafellar \cite{SLSQP5} have introduced the use of penalty functions in the nonlinear programming to control the step size.

The {\bf Critical Line Algorithm (CLA)} is an efficient alternative to the quadratic optimizer for mean-variance model, as it is specifically designed for inequality portfolio optimization. It was already originally introduced in the Markowitz Portfolio Selection paper \cite{Markowitz}, and its computational implementation has become increasingly popular \cite{CLA2,CLA3}. CLA also solves constrained problems with conditions in inequalities, but unlike SLS, it divides a constrained problem into series of unconstrained sub-problems by invoking the concept of turning point. A turning point is a constrained minimum variance portfolio whose vicinity contains other constrained minimum variance portfolios of different free assets.

Similar to quadratic programming, an initial solution is required on the constrained minimum variance frontier. To construct the initial solution, assets are ranked with respect to their expected returns. Then, one increases the weight of the first asset of the highest expected returns, $w_1$, from a defined lower bound $l_1 = 0$ to an upper bound $u_1$ if $w_1 \leq 1$. Subsequently, the following assets have their weights increased until $\sum_i{w_i}=1$. Typically, the weights of the first  and the last few assets are set to the upper and lower bound which are called bounded assets, while only one in the middle has its weight between bounds and referred as the free asset. The free weight is expressed as:
\begin{equation}
 {w_f}=1-\sum_{i \in \mathds{U}}{w_i}-\sum_{i \in \mathds{L}}{w_i}
\end{equation}
where $\mathds{U}$ and $\mathds{L}$ represents two sets of upper and lower bounded weights. Then in the following iterations, by decreasing the Lagrange multiplier for the constraint on expected portfolio return, $\lambda$ to move to the next lower turning point, two cases need to be considered to compute $\mathbf{W}$. A formally free asset moves to its bound, or vice versa, a bounded asset wants to become free. In both situations, the maximum threshold $\lambda_{in}$ and $\lambda_{out}$ for the former and the later will be found. Subsequently, the larger one characterises the new turning point, and the asset is moved accordingly, and weights are re-assigned. As the free and bounded assets do not interchange between turning points, the constrained solution between two turning points is in fact the solution of unconstrained optimization on only the free assets. Therefore, the constrained problem reduces to solving the unconstrained problem on the free assets. When no new threshold can be found, the lowest turning point is said to be reached and the algorithm is terminated for the optimized $\mathbf{W}$.

%

\end{appendices}
\newpage


\newpage

\end{document}